\documentclass[natbib]{svjour3}

\usepackage{setspace}
\usepackage{amsmath}
\usepackage{amssymb}
\usepackage{graphicx}

\title{Transition probabilities for general birth-death processes with applications in ecology, genetics, and evolution}

\titlerunning{Birth-death processes}

\author{Forrest W. Crawford \and Marc A. Suchard}

\institute{Forrest W. Crawford \at
  Department of Biomathematics,
  University of California Los Angeles
  Los Angeles, CA 90095-1766 USA
  \email{fcrawford@ucla.edu}
      \and
  Marc A. Suchard \at
  Departments of Biomathematics, Biostatistics and Human Genetics,
  University of California Los Angeles
  Los Angeles, CA 90095-1766 USA
  \email{msuchard@ucla.edu}
}

\date{Typeset on \today}

\smartqed

\newcommand{\od}[2]{\frac{\text{d} #1}{\text{d} #2}}

\newcommand{\dx}[1]{\ \text{d} #1}
\newcommand{\E}{\mathbb{E}}

\renewcommand{\Re}{\text{Re}}
\renewcommand{\Im}{\text{Im}}

\begin{document}

\maketitle

\begin{abstract} 
\noindent A birth-death process is a continuous-time Markov chain that counts the number of particles in a system over time.  In the general process with $n$ current particles, a new particle is born with instantaneous rate $\lambda_n$ and a particle dies with instantaneous rate $\mu_n$.  Currently no robust and efficient method exists to evaluate the finite-time transition probabilities in a general birth-death process with arbitrary birth and death rates.  In this paper, we first revisit the theory of continued fractions to obtain expressions for the Laplace transforms of these transition probabilities and make explicit an important derivation connecting transition probabilities and continued fractions.  We then develop an efficient algorithm for computing these probabilities that analyzes the error associated with approximations in the method.  We demonstrate that this error-controlled method agrees with known solutions and outperforms previous approaches to computing these probabilities.  Finally, we apply our novel method to several important problems in ecology, evolution, and genetics.


\PACS{PACS code1 \and PACS code2 \and more}
\end{abstract}


\section{Introduction}

Birth-death processes (BDPs) have a rich history in probabilistic modeling, including applications in ecology, genetics, and evolution \citep{Thorne1991Evolutionary,Krone1997Ancestral,Novozhilov2006Biological}.  Traditionally, BDPs have been used to model the number of organisms or particles in a system, each of which reproduce and die in continuous time.  A general BDP is a continuous-time Markov chain on the non-negative integers in which instantaneous transitions from state $n\geq 0$ to either $n+1$ or $n-1$ are possible.  These transitions are called ``births'' and ``deaths''.  Starting at state $n$, jumps to $n+1$ occur with instantaneous rate $\lambda_n$ and jumps to $n-1$ with instantaneous rate $\mu_n$.  The simplest BDP has linear rates $\lambda_n = n\lambda$ and $\mu_n = n\mu$ with no state-independent terms \citep{Kendall1948Generalized,Feller1971Introduction}.  This model is the most widely-used BDP since there exist closed-form expressions for its transition probabilities \citep{Bailey1964Elements,Novozhilov2006Biological}.  Many applications of BDPs require convenient methods for computing the probability $P_{m,n}(t)$ that the system moves from state $m$ to state $n$ in finite time $t\geq 0$.  These probabilities exhibit their usefulness in many modeling applications since the probabilities do not depend on the possibly unobserved path taken by the process from $m$ to $n$ and hence make possible analyses of discretely sampled or partially observed processes.  Despite the relative simplicity of specifying the rates of a general BDP, it can be remarkably difficult to find closed-form solutions for the transition probabilities even for simple models \citep{Renshaw1993Modelling,Mederer2003Transient,Novozhilov2006Biological}.

In a pioneering series of papers, \citeauthor{Karlin1957Differential} develop a formal theory of general BDPs that expresses their transition probabilities in terms of a sequence of orthogonal polynomials and a spectral measure \citep{Karlin1957Classification,Karlin1957Differential,Karlin1958Many}.  While the work of \citeauthor{Karlin1957Differential} yields valuable theoretical insights regarding the existence of unique solutions and properties of recurrence and transience for a given process, there remains no clear recipe for determining the orthogonal polynomials and measure corresponding to an arbitrary set of birth and death rates.  Additionally, even when the polynomials and measure are known, the transition probabilities may not have an analytic representation or a convenient computational form.  

Possibly due to the difficulty of finding computationally useful formulas for transition probabilities in general BDPs, many applied researchers resort to easier analyses using moments, first passage times, equilibrium probabilities, and other tractable quantities of interest.  Referring to the system of Kolmogorov forward differential equations for transition probabilities that we give below, \citet[page 73]{Novozhilov2006Biological} write,
\begin{quote}
``The problem with exact solutions of system (1) is that, in many cases, the expressions for the state probabilities, although explicit, are intractable for analysis and include special polynomials. In such cases, it may be sensible to solve more modest problems concerning the birth-and-death process under consideration, without the knowledge of the time-dependent behavior of state probabilities $p_n(t)$.''
\end{quote}
Indeed, closed-form analytic expressions for transition probabilities of general BDPs are only known for a few types of processes. Some examples include constant birth and death rates \citep{Bailey1964Elements}, zero birth or death rates (pure-death and pure-birth) \citep{Yule1925Mathematical,Taylor1998Introduction}, and certain linear rates \citep{Karlin1958Linear}.  As a seemingly straightforward example, in the BDP with linear birth and death rates $\lambda_n=n\lambda+\nu$ and $\mu_n=n\mu+\gamma$ including state-independent terms, \citet{Ismail1988Linear} offer the orthogonal polynomials and associated measure, but still no closed form is available for the transition probabilities.  

Despite the difficulty in obtaining analytic expressions, several authors have made progress in approximate numerical methods for solution of transition probabilities in general BDPs.  \citet{Murphy1975Some} develop an appealing numerical method for the transition probabilities based on a continued fraction representation of Laplace-transformed transition probabilities.  They invert these transformed probabilities by first truncating the continued fraction.  Several other authors give similar expressions derived from truncation of the state space \citep{Grassmann1977Transienta,Grassmann1977Transientb,Rosenlund1978Transition,Sharma1988Multi,Mohanty1993Transient}.  However, \citet{Klar2010Zipf} find that methods based on continued fraction truncation and then subsequent analytical transformation can suffer from instability.  As an alternative, \citet{Parthasarathy2005Exact} express the infinite continued fraction representation given by \citeauthor{Murphy1975Some} as a power series.  Unfortunately, the small radius of convergence of this series makes it less useful for numerical computation.

We also note that for general BDPs that take values on a finite state space (usually $n\in\{0,1,\ldots,N\}$), it is possible to write a finite-dimensional stochastic transition rate matrix and solve for the matrix of transition probabilities.  If the rate matrix is diagonalizable, computation of transition probabilities in this manner can be computationally straightforward.  To illustrate, let $Q$ be a finite-dimensional stochastic rate matrix with $Q=U\Lambda U^{-1}$ where $U$ is an orthogonal matrix and $\Lambda$ is diagonal. The matrix of transition probabilities $P$ satisfies the matrix differential equation $P' = PQ$ with initial condition $P(0)=I$.  The solution is $P(t) = \exp[Qt] = U\ \text{diag}( e^{z_1 t}, e^{z_2 t}, \dots, e^{z_N t})\ U^{-1}$, where $z_1,\ldots,z_N$ are the eigenvalues of $Q$.  However, it is possible to specify reasonable rate parameters in a general BDP that satisfy requirements for the existence of a unique solution, but do not result in a diagonalizable rate matrix.  Also, if the state space over which the BDP takes values is large, numerical eigendecomposition of $Q$ may be computationally expensive and could introduce serious roundoff errors.  

To our knowledge, no robust computational method currently exists for finding the finite-time transition probabilities of general BDPs with arbitrary rates.  Such a technique would allow rapid development of rich and sophisticated ecological, genetic, and evolutionary models.  Additionally, in statistical applications, transition probabilities can serve as observed data likelihoods, and are thus often useful in estimating transition rate parameters from partially observed BDPs.  We believe more sophisticated BDPs can be very useful for applied researchers.  In spite of the numerical difficulties presented by approximant methods, we are surprised that continued fraction methods like that of \citet{Murphy1975Some} are not more widely explored.  This may be due to omission of important details in their derivation of continued fraction expressions for the Laplace transform of the transition probabilities.  

In this paper, we build on continued fraction expressions for the Laplace transforms of the transition probabilities of a general BDP using techniques similar to those introduced by \citeauthor{Murphy1975Some}, and we fill in the missing details in the proof of this representation.  We then apply the Laplace inversion formulae of \citet{Abate1992Fourier,Abate1992Numerical} to obtain an efficient and robust method for computation of transition probabilities in general BDPs.  Our method relies on three observations: 1) it is possible to find exact expressions for Laplace transforms of the transition probabilities of a general BDP using continued fractions \citep{Murphy1975Some}; 2) evaluation of continued fractions is typically very fast, requires far fewer evaluations than equivalent power series, and there exist robust algorithms for evaluating them efficiently \citep{Bankier1942Numerical,Wall1948Analytic,Blanch1964Numerical,Lorentzen1992Continued,Craviotto1993Survey,Abate1999Computing,Cuyt2008Handbook}; and 3) recovery of probability distributions by Laplace inversion using a Riemann sum approximation is often more computationally stable than analytical methods of inversion \citep{Abate1992Fourier,Abate1992Numerical,Abate1995Numerical}.  Finally, we demonstrate the advantages of our error-controlled method through its application to several birth-death models in ecology, genetics, and evolution whose solution remains unavailable by other means.

\section{Transition probabilities}

\subsection{Background}

A general birth-death process is a continuous-time Markov process $\mathcal{X} = \{X(t), t\geq 0\}$ counting the number of arbitrarily defined ``particles'' in existence at time $t\geq 0$, with $X(0) = m \geq 0$.  To characterize the process, we define non-negative instantaneous birth rates $\lambda_n$  and death rates $\mu_n$ for $n\geq 0$, with $\mu_0 = 0$ and transition probabilities $P_{m,n}(t) = \Pr(X(t) = n \mid X(0) = m)$.  While $\lambda_n$ and $\mu_n$ are time-homogeneous constants, they may depend on $n$.  We refer to the classical linear BDP in which $\lambda_n = n\lambda$ and $\mu_n = n\mu$ as the ``simple birth-death process'' \citep{Kendall1948Generalized,Feller1971Introduction}.  The general BDP transition probabilities satisfy the infinite system of ordinary differential equations
\begin{equation} 
\label{eq:odes}
\begin{split}
	\od{P_{m,0}(t)}{t} &= \mu_1 P_{m,1}(t) -\lambda_0 P_{m,0}(t) \text{, and} \\
	\od{P_{m,n}(t)}{t} &= \lambda_{n-1}P_{m,n-1}(t) + \mu_{n+1}P_{m,n+1}(t) - (\lambda_n + \mu_n)P_{m,n}(t) \text{ for $n\geq 1$,}
\end{split}
\end{equation}
with boundary conditions $P_{m,m}(0) = 1$ and $P_{m,n}(0) = 0$ for $n \neq m$ \citep{Feller1971Introduction}. 

\citet{Karlin1957Differential} show that for arbitrary starting state $m$, transition probabilities can be represented in the form
\begin{equation}
	P_{m,n}(t) = \pi_n \int_0^\infty e^{-xt} Q_m(x) Q_n(x) \psi(\text{d}x),
	\label{eq:integral}
\end{equation}
where $\pi_0 = 1$ and $\pi_n = (\lambda_0\cdots\lambda_{n-1})/(\mu_1\cdots\mu_n )$ for $n \geq 1$.  Here, $\{Q_n(x)\}$ is a sequence of polynomials satisfying the three-term recurrence relation
\begin{equation}
\begin{split}
  \lambda_0 Q_{1}(x) &= \lambda_0 + \mu_0 - x \text{, and} \\
  \lambda_n Q_{n+1}(x) &= (\lambda_n + \mu_n - x)Q_n(x) - \mu_n Q_{n-1}(x), 
\end{split}
\label{eq:orthogpoly}
\end{equation}
and $\psi$ is the spectral measure of $\mathcal{X}$ with respect to which the polynomials $\{Q_n(x)\}$ are orthogonal.  The system \eqref{eq:odes} has a unique solution if and only if
\begin{equation}
  \sum_{k=0}^\infty \left( \pi_k + \frac{1}{\lambda_k \pi_k} \right) = \infty. 
\label{eq:moment}
\end{equation}
In what follows, we assume that the rate parameters $\{\lambda_n\}$ and $\{\mu_n\}$ satisfy \eqref{eq:moment}.  Closed-form solutions to \eqref{eq:odes} are available for a surprisingly small number of choices of $\{\lambda_n\}$ and $\{\mu_n\}$.  We therefore need another approach to find useful formulae for computation of the transition probabilities.

\subsection{Continued fraction representation of Laplace transform}

To find an expression that is useful for computing $P_{m,n}(t)$ for an arbitrary general BDP, a fruitful approach is often to Laplace transform each equation of the system \eqref{eq:odes} and form a recurrence relationship relating back to the Laplace transform of $P_{m,n}(t)$. We base our presentation on that of \citet{Murphy1975Some}.  Denote the Laplace transform of $P_{n,m}(t)$ as
\begin{equation}
 f_{m,n}(s) = \mathcal{L}\left[P_{m,n}(t)\right](s) = \int_0^\infty e^{-st} P_{m,n}(t) \dx{t} .
\end{equation}
Applying the Laplace transform to \eqref{eq:odes}, with the starting state $m=0$, we arrive at
\begin{equation}
\begin{split}
  s f_{0,0}(s) - P_{0,0}(0) &=  \mu_1 f_{0,1}(s) - \lambda_0 f_{0,0}(s) \text{, and} \\
  s f_{0,n}(s) - P_{0,n}(0) &=  \lambda_{n-1}f_{0,n-1}(s) + \mu_{n+1} f_{0,n+1}(s) - (\lambda_n + \mu_n)f_{0,n}(s) 
 \end{split}
\label{eq:recur0}
\end{equation}
for $n\geq 1$.  Rearranging and recalling that $P_{0,0}(0) = 1$ and $P_{0,n}(0) = 0$ for $n\geq 1$, we simplify \eqref{eq:recur0} to 
\begin{equation}
 \begin{split}
	 f_{0,1}(s) &=  \frac{1}{\mu_1}\big[ (s+\lambda_0) f_{0,0}(s) - 1 \big], \text{and} \\
	 f_{0,n}(s) &=  \frac{1}{\mu_n}\bigg[ (s+\lambda_{n-1} + \mu_{n-1}) f_{0,n-1}(s) - \lambda_{n-2}f_{0,n-2}(s)\bigg] \text{for $n\geq 2$}. 
\end{split}
 \label{eq:recur1}
\end{equation}
Some rearranging of \eqref{eq:recur1} yields the forward system of recurrence relations
\begin{equation}
	\begin{split}
		f_{0,0}(s) &= \frac{1}{s + \lambda_0 - \mu_1 \left(\frac{f_{0,1}(s)}{f_{0,0}(s)}\right) } \text{, and} \\
		\frac{f_{0,n}(s)}{f_{0,n-1}(s)} &= \frac{\lambda_{n-1}}{s + \mu_n + \lambda_{n} - \mu_{n+1} \left(\frac{f_{0,n+1}(s)}{f_{0,n}(s)}\right)}. 
 \end{split}
 \label{eq:recur2}
\end{equation}
Then combining these expressions, we arrive at the generalized continued fraction 
\begin{equation}
  f_{0,0}(s) = \cfrac{1}{s+\lambda_0 - \cfrac{\lambda_0 \mu_1}{s+\lambda_1+\mu_1 - \cfrac{\lambda_1 \mu_2}{s+\lambda_2+\mu_2 - \cdots}}}.
  \label{eq:cfrac1}
\end{equation}
This is an exact expression for the Laplace transform of the transition probability $P_{0,0}(t)$.  Let the partial numerators in \eqref{eq:cfrac1} be $a_1 = 1$ and $a_n = -\lambda_{n-2}\mu_{n-1}$, and the partial denominators $b_1 = s+\lambda_0$ and $b_n = s+\lambda_{n-1}+\mu_{n-1}$ for $n\geq 2$.  Then \eqref{eq:cfrac1} becomes
\begin{equation}
  f_{0,0}(s) = \cfrac{a_1}{b_1 + \cfrac{a_2}{b_2 + \cfrac{a_3}{b_3 + \cdots}}}.
  \label{eq:cfrac2}
\end{equation}
To express \eqref{eq:cfrac2} in more typographically economical notation, we write
\begin{equation}
  f_{0,0}(s) = \frac{a_1}{b_1+} \frac{a_2}{b_2+} \frac{a_3}{b_3+} \cdots .
 \label{eq:cfrac3}
\end{equation}
We denote the $k$th convergent (approximant) of $f_{0,0}(s)$ as
\begin{equation}
 f_{0,0}^{(k)}(s) = \frac{a_1}{b_1+} \frac{a_2}{b_2+} \cdots \frac{a_k}{b_k} = \frac{A_k(s)}{B_k(s)}.
\label{eq:convergent}
\end{equation}
There are deep connections between the orthogonal polynomial representation \eqref{eq:orthogpoly}, Laplace transforms \eqref{eq:recur1}, and continued fractions of the form \eqref{eq:cfrac1} that are beyond the scope of this paper \citep{Karlin1957Differential,Bordes1983Application,Guillemin1999Excursions}.  Interestingly, \citet{Flajolet2000Formal} demonstrate a close relationship between the Laplace transforms of transition probabilities and state paths of the underlying Markov chain.

Before stating a theorem supporting this representation, we give two lemmas that will be useful in what follows.
\newtheorem{ctr:wallis}{Lemma}
\begin{ctr:wallis}
 \label{lem:wallis}
Both the numerator $A_k$ and denominator $B_k$ of \eqref{eq:convergent} satisfy the same recurrence, due to \citet{Wallis1695Opera}:
\begin{equation}
 \begin{split}
  A_k &= b_k A_{k-1} + a_k A_{k-2} \text{, and} \\
  B_k &= b_k B_{k-1} + a_k B_{k-2} ,
\end{split}
\end{equation}
with $A_0 = 0$, $A_1 = a_1$, $B_0 = 1$, and $B_1 = b_1$.  
\end{ctr:wallis}
\newtheorem{ctr:det}[ctr:wallis]{Lemma}
\begin{ctr:det}
	\label{lem:det}
	By repeated application of Lemma \ref{lem:wallis}, we arrive at the determinant formula
\begin{equation}
	\begin{split}
	A_k B_{k-1} - A_{k-1}B_k &= (b_k A_{k-1} + a_k A_{k-2})B_{k-1} - A_{k-1}(b_k B_{k-1} + a_k B_{k-2}) \\
		&= -a_k(A_{k-1}B_{k-2} - A_{k-2}B_{k-1})  \\
		&= (-1)^{k-1} \prod_{i=1}^k a_i .
	\end{split}
	\label{eq:det}
\end{equation}
\end{ctr:det}
Now we state and prove a theorem giving expressions for the Laplace transform of $P_{m,n}(t)$.  Although \citet{Murphy1975Some} first report this result, they do not provide a detailed derivation in their paper.  
\newtheorem{thm:cf}{Theorem}
\begin{thm:cf}
	\label{thm:cf}
	The Laplace transform of the transition probability $P_{m,n}(t)$ is given by 
\begin{equation}
	f_{m,n}(s) = \begin{cases} 
		\displaystyle\left(\prod_{j=n+1}^m \mu_j\right)\frac{B_n(s)}{B_{m+1}(s)+} \frac{B_m(s) a_{m+2}}{b_{m+2}+} \frac{a_{m+3}}{b_{m+3}+}\cdots  & \text{for $n\leq m$}, \\
		& \\
	\displaystyle\left(\prod_{j=m}^{n-1} \lambda_j\right) \frac{B_m(s)}{B_{n+1}(s)+} \frac{B_n(s) a_{n+2}}{b_{n+2}+} \frac{a_{n+3}}{b_{n+3}+}\cdots  & \text{for $m \leq n$,}
\end{cases}
	\label{eq:fmnfull}
\end{equation}
where $a_n$, $b_n$, and $B_n$ are as defined above.
\end{thm:cf}

\begin{proof} To simplify notation, we sometimes omit the dependence of $f_k$, $A_k$, and $B_k$ on the Laplace variable $s$.  Suppose the process starts at $X(0)=m$.  We can re-write the Laplace-transformed equations \eqref{eq:recur0} with $P_{m,m}(0)=1$ and $P_{m,n}(0)=0$ for all $n\neq m$ as
\begin{subequations}
	\label{eq:fmnlts}
	\begin{align}
		sf_{m,0}(s) - \delta_{m0} &= \mu_1 f_{m,1}(s) - \lambda_0 f_{m,0}(s),\label{eq:fm0} \\
		sf_{m,n}(s) - \delta_{mn} &= \lambda_{n-1}f_{m,n-1}(s) + \mu_{n+1}f_{m,n+1}(s) - (\lambda_n + \mu_n)f_{m,n}(s),\label{eq:fmn}
	\end{align}
	\label{eq:grecur}
\end{subequations}
where $\delta_{mn}=1$ if $m=n$ and zero otherwise. We first derive the expression for $n\leq m$. If $m=0$, $f_{0,0}(s)$ is given by \eqref{eq:cfrac3}, so we assume in what follows when $n\leq m$, that $m\geq 1$.  Rearranging \eqref{eq:fm0}, we see that since $B_0=1$ and $s+\lambda_0=b_1=B_1$,
\begin{equation}
f_{m,0} = \frac{B_0}{B_1}\mu_1 f_{m,1}. 
	\label{eq:fm0recur}
\end{equation}
Now, to show the general case by induction, assume that for $n\leq m$,
\begin{equation}
 f_{m,n-1} = \frac{B_{n-1}}{B_n}\mu_n f_{m,n}.
	\label{eq:fmn1}
\end{equation}
Substituting \eqref{eq:fmn1} into \eqref{eq:fmn} when $n<m$, we have
\begin{equation}
 b_{n+1}f_{m,n} = \lambda_{n-1}\frac{B_{n-1}}{B_n}\mu_n f_{m,n} + \mu_{n+1}f_{m,n+1}
\end{equation}
\begin{equation}
 \left(b_{n+1} + a_{n+1}\frac{B_{n-1}}{B_n} \right)f_{m,n} = \mu_{n+1}f_{m,n+1}
\end{equation}
\begin{equation}
 f_{m,n} = \frac{B_{n}}{B_{n+1}}\mu_{n+1}f_{m,n+1}
\end{equation}
and so \eqref{eq:fmn1} is true for any $n<m$.  Letting $n=m$, we have by \eqref{eq:fmn1} and \eqref{eq:fmn},
\begin{equation}
 b_{m+1}f_{m,m} = 1 + \lambda_{m-1}\left( \frac{B_{m-1}}{B_m}\mu_m f_{m,m}\right) + \mu_{m+1}f_{m,m+1}.
\end{equation}
Recalling that $s+\lambda_m+\mu_m = b_{m+1}$ and using Lemma \ref{lem:wallis},
\begin{equation}
 \mu_{m+1}f_{m,m+1} = 1 - \frac{B_{m+1}}{B_m} f_{m,m}.
	\label{eq:fmmrecur1}
\end{equation}
Rearranging the previous equation, we find that 
\begin{equation}
	f_{m,m} = \frac{1}{\frac{B_{m+1}}{B_m} + \mu_{m+1}\frac{f_{m,m+1}}{f_{m,m}}}.
	\label{eq:fmmrecur2}
\end{equation}
Likewise, we can write \eqref{eq:fmn} as a continued fraction recurrence:
\begin{equation}
 \frac{f_{m,n}}{f_{m,n-1}} = \frac{\lambda_{n-1}}{s+\mu_n+\lambda_n + \mu_{n+1}\frac{f_{m,n+1}}{f_{m,n}}}.
	\label{eq:fmnfmn1}
\end{equation}
Then plugging \eqref{eq:fmnfmn1} into \eqref{eq:fmmrecur2} and iterating, we obtain the continued fraction for $f_{m,m}$:
\begin{equation}
	\begin{split}
		f_{m,m} &= \frac{1}{\frac{B_{m+1}}{B_m}+} \frac{a_{m+2}}{b_{m+2}+} \frac{a_{m+3}}{b_{m+3}+}\cdots \\
		&= \frac{B_m}{B_{m+1}+} \frac{B_m a_{m+2}}{b_{m+2}+} \frac{a_{m+3}}{b_{m+3}+}\cdots .
\end{split}
	\label{eq:fmmcf}
\end{equation}
This is an exact formula for the Laplace transform of $P_{m,m}(t)$, and proves the case $m=n$.  For $n\leq m$, we iterate \eqref{eq:fmn1} to get
\begin{equation}
	\begin{split}
  f_{m,n} &= \frac{B_n}{B_{n+1}}\mu_{n+1}f_{m,n+1} \\
	&= \frac{B_n}{B_{n+1}} \frac{B_{n+1}}{B_{n+2}} \mu_{n+1}\mu_{n+2} f_{m,n+2} \\
	&= \frac{B_n}{B_{n+1}} \frac{B_{n+1}}{B_{n+2}}\cdots\frac{B_{m-1}}{B_m} \mu_{n+1}\mu_{n+2}\cdots\mu_m f_{m,m} \\
	&= \left(\prod_{j=n+1}^m \mu_j\right) \frac{B_n}{B_m} f_{m,m} .
\end{split}
	\label{eq:fmnless}
\end{equation}
Substituting \eqref{eq:fmmcf} for $f_{m,m}$ completes the proof for $n\leq m$.

To find the formula for $f_{m,n}$ when $n>m$, we adopt a similar approach.  From \eqref{eq:fmmrecur2} we arrive at 
\begin{equation}
B_{m+1}f_{m,m} = B_m - B_m\mu_{m+1}f_{m,m+1}.
	\label{eq:fmmrecurup}
\end{equation}
We proceed inductively. Assume that for $n>m$,
\begin{equation}
 B_{n+1} f_{m,n} = \left(\prod_{j=m}^{n-1} \lambda_j\right) B_m + \mu_{n+1} B_{n} f_{m,n+1}.
	\label{eq:fmn1recur}
\end{equation}
From \eqref{eq:fmn}, we have 
\begin{equation} 
b_{n+2}f_{m,n+1} = \lambda_nf_{m,n} + \mu_{n+2}f_{m,n+2} .
\end{equation}
Solving for $f_{m,n}$ in \eqref{eq:fmn1recur} and plugging this into the above equation, we have
\begin{equation}
 b_{n+2}f_{m,n+1} = \lambda_n\left(\prod_{j=m}^{n-1} \lambda_j\right) \frac{B_m}{B_{n+1}} + \lambda_n\mu_{n+1}\frac{B_n}{B_{n+1}}f_{m,n+1} + \mu_{n+2}f_{m,n+2}.
 \end{equation}
Recalling that $-\lambda_n\mu_{n+1} = a_{n+2}$, 
\begin{equation} 
\left( b_{n+2}B_{n+1} + a_{n+2} B_m \right)f_{m,n+1} = \left(\prod_{j=m}^n\lambda_j\right) B_n + \mu_{n+2} B_{n+1} f_{m,n+2}, 
  \end{equation}
and by Lemma \ref{lem:wallis},
\begin{equation} 
B_{n+2} f_{m,n+1} = \left(\prod_{j=m}^n\lambda_j\right) B_m + \mu_{m+2} B_{n+1} f_{m,m+2}. 
\end{equation}
This establishes the recurrence \eqref{eq:fmn1recur}.  Then for any $n\geq m$, we can rearrange \eqref{eq:fmn1recur} to obtain
\begin{equation}
f_{m,n} = \left(\prod_{j=m}^{n-1}\lambda_j\right) \frac{B_m}{B_{n+1} - B_n\mu_{n+1}\frac{f_{m,n+1}}{f_{m,n}}}. 
\end{equation}
This completes the proof.  \qed
\end{proof}


\subsection{Obtaining transition probabilities}

\citet{Murphy1975Some} find transition probabilities by truncating \eqref{eq:fmnfull} at a pre-specified depth, forming a partial fractions sum, and inverse transforming.  \citet{Parthasarathy2005Exact} give a series solution for transition probabilities based on an equivalence between continued fractions like \eqref{eq:fmnfull} and power series.  However, both of these approaches suffer from serious drawbacks, as we explore in detail in the Appendix.

We instead seek an efficient and robust numerical method for evaluating and inverting \eqref{eq:fmnfull}.  We first note that continued fractions typically converge rapidly, and in our experience, evaluation of \eqref{eq:fmnfull} is very fast and stable using the Lentz algorithm and its subsequent improvements \citep{Lentz1976Generating,Thompson1986Coulomb,Press2007Numerical}.  We therefore invert \eqref{eq:fmnfull} numerically by a summation formula.

To do this, we treat the continued fraction representation \eqref{eq:fmnfull} of the Laplace transform of $P_{m,n}(t)$ as an unknown but computable function of the complex Laplace variable $s$.  We base our presentation on that of \citet{Abate1992Fourier}.  If $\epsilon$ is a positive real number such that all singularities of $f_{m,n}(s)$ lie to the left of $\epsilon$ in the complex plane, the inverse Laplace transform of $f_{m,n}(s)$ is given by the Bromwich integral
\begin{equation}
  P_{m,n}(t) = \mathcal{L}^{-1}\left(f_{m,n}(s)\right) = \frac{1}{2\pi i} \int_{\epsilon-i\infty}^{\epsilon+i\infty} e^{st}f_{m,n}(s)\dx{s} .
\end{equation}
Letting $s=\epsilon+iu$,
\begin{equation}
\begin{split}
 P_{m,n}(t) &= \frac{1}{2\pi} \int_{-\infty}^{\infty} e^{(\epsilon+iu)t} f_{m,n}(\epsilon+iu)\dx{u} \\
   &= \frac{e^{\epsilon t}}{2\pi} \int_{-\infty}^{\infty} \big[\cos(ut) + i\sin(ut) \big] f_{m,n}(\epsilon+iu)\dx{u} \\
   &= \frac{e^{\epsilon t}}{2\pi} \Bigg[ \int_{-\infty}^{\infty} \Big[\Re\big(f_{m,n}(\epsilon+iu)\big)\cos(ut) - \Im\big(f_{m,n}(\epsilon+iu)\big)\sin(ut) \Big] \dx{u} \\
    &\quad +i\int_{-\infty}^\infty \Big[\Im\big(f_{m,n}(\epsilon+iu)\big)\cos(ut) + \Re\big(f_{m,n}(\epsilon+iu)\big)\sin(ut)\Big] \dx{u} \Bigg] ,
\end{split}
\label{eq:inversion1}
\end{equation}
but $P_{m,n}(t)$ is real-valued, so the imaginary part of the last equality in \eqref{eq:inversion1} is zero.  Then 
\begin{equation}
 P_{m,n}(t) = \frac{e^{\epsilon t}}{2\pi} \int_{-\infty}^{\infty} \Big[\Re\big(f_{m,n}(\epsilon+iu)\big)\cos(ut) - \Im\big(f_{m,n}(\epsilon+iu)\big)\sin(ut)\Big]  \dx{u} .
\label{eq:inversion2}
\end{equation}
But since $P_{m,n}(t)=0$ for $t<0$, we also have that 
\begin{equation}
 \int_{-\infty}^\infty \Big[\Re\big(f_{m,n}(\epsilon+iu)\big)\cos(ut) + \Im\big(f_{m,n}(\epsilon+iu)\big)\sin(ut)\Big]  \dx{u} = 0 .
\label{eq:neg}
\end{equation}
Then applying \eqref{eq:neg} to \eqref{eq:inversion2}, we obtain
\begin{equation}
P_{m,n}(t) = \frac{e^{\epsilon t}}{\pi} \int_{-\infty}^\infty \Re\big(f_{m,n}(\epsilon+iu)\big)\cos(ut) \dx{u}. 
 \label{eq:inversion3}
\end{equation}
Finally, we note that since 
\begin{equation}
\Re\big(f(\epsilon-iu)\big) = \int_0^\infty e^{-\epsilon t} \cos(ut) P_{m,n}(t)\dx{t} = \Re\big(f(\epsilon+iu)\big) ,
\label{eq:feven}
\end{equation}
it must be the case that $\Re\big(f_{m,n}(\epsilon+iu)\big)$ is even in $u$ for every $\epsilon$.  Therefore, 
\begin{equation}
   P_{m,n}(t) = \frac{2e^{\epsilon t}}{\pi} \int_0^\infty \Re\big(f_{m,n}(\epsilon+iu)\big)\cos(ut) \dx{u}.
\label{eq:inversion4}
\end{equation}

Following \citet{Abate1992Fourier}, we approximate the integral above by a discrete Riemann sum via the trapezoidal rule with step size $h$:
\begin{equation}
	\begin{split}
		P_{m,n}(t) &\approx \frac{he^{\epsilon t}}{\pi} \Re\left(f_{m,n}(\epsilon )\right) + \frac{2he^{\epsilon t}}{\pi} \sum_{k=1}^\infty \Re\left(f_{m,n}(\epsilon +ikh)\right) \cos(kht) \\
		&= \frac{e^{A/2}}{2t} \Re\left(f_{m,n}\left(\frac{A}{2t}\right)\right) + \frac{e^{A/2}}{t} \sum_{k=1}^\infty (-1)^k \Re\left(f_{m,n}\left(\frac{A+2k\pi i}{2t}\right)\right),  \\
	\end{split}
	\label{eq:quad}
\end{equation}
where the second line is obtained by setting $h = \pi/(2t)$ and $\epsilon = A/(2t)$; this change of variables eliminates the cosine term.  

\subsection{Numerical considerations}

While \eqref{eq:quad} presents a method for numerical solution of the transition probabilities $P_{m,n}(t)$ for a BDP with arbitrary birth and death rates, it is not yet an algorithm for reliable evaluation of these probabilities.  In order to develop a reliable numerical method, we must: 1) characterize the error introduced by discretization of the integral in \eqref{eq:inversion4}; 2) determine a suitable method to evaluate this nearly alternating sum while controlling the error; and 3) accurately and rapidly evaluate the infinite continued fraction in \eqref{eq:fmnfull}.

\citeauthor{Abate1992Fourier} show that the discretization error that arises in \eqref{eq:quad} is
\begin{equation}
 e_d = \sum_{k=1}^\infty e^{-kA} P_{m,n}\big( (2k+1)t \big), 
 \end{equation}
and when $P_{m,n}(t)\leq 1$,
\begin{equation}
e_d \leq \sum_{k=1}^\infty e^{-kA} = \frac{e^{-A}}{1-e^{-A}} \approx e^{-A}, 
    \end{equation}
when $e^{-A}$ is small.  Then to obtain $e_d \leq 10^{-\gamma}$, we set $A = \gamma\log(10)$.  As \citeauthor{Abate1992Fourier} point out, the terms of the series \eqref{eq:quad} alternate in sign when
\begin{equation}
\Re\left(f_{m,n}\left(\frac{A+2k\pi i}{2t}\right)\right)  
  \end{equation}
has constant sign.  This suggests that a series acceleration method may be helpful in keeping the terms of the sum manageable and avoiding roundoff error due to summands of alternating sign.  We opt to use the Levin transform for this purpose \citep{Levin1973Non,Press2007Numerical,Numerical2007Derivation}.

Evaluation of rational approximations to continued fractions by repeated application of Lemma \ref{lem:wallis} is appealing, but suffers from roundoff error when denominators are small \citep{Press2007Numerical}.  To evaluate the infinite continued fraction in the summand of \eqref{eq:quad}, we use the modified Lentz method \citep{Lentz1976Generating,Thompson1986Coulomb,Press2007Numerical}. To demonstrate, suppose we wish to approximate the value of $f_{0,0}(s)$, given by \eqref{eq:cfrac1} by truncating at depth $k$.  Then 
\begin{equation} 
f_{0,0}^{(k)}(s) = \frac{A_k(s)}{B_k(s)} 
\end{equation}
is the $k$th rational approximant to the infinite continued fraction $f_{0,0}(s)$.  In the modified Lentz method, we stabilize the computation by finding the ratios
\begin{equation} 
C_k = \frac{A_k}{A_{k-1}} \quad \text{and}\quad D_k = \frac{B_{k-1}}{B_k} 
\end{equation}
so that $f_{0,0}^{(k)}$ can be found iteratively by 
\begin{equation} 
f_{0,0}^{(k)} = f_{0,0}^{(k-1)} C_k D_k. 
\end{equation}
Using Lemma \ref{lem:wallis}, we can iteratively compute $C_k$ and $D_k$ via the updates
\begin{equation} 
C_k = b_k + \frac{a_k}{C_{k-1}} \quad \text{and}\quad D_k = \frac{1}{b_k + a_k D_{k-1}}. 
\end{equation}

In practice, we must evaluate the continued fraction to only a finite depth, but we must evaluate to a depth sufficient to control the error. Suppose we wish to evaluate the infinite continued fraction $f_{0,0}(s)$ given by \eqref{eq:cfrac1} at some complex number $s$.  Intuitively, we wish to terminate the Lentz algorithm when the difference between successive convergents is small.  However, it is not immediately clear how the difference between convergents $f_{0,0}^{(k)}(s)-f_{0,0}^{(k-1)}(s)$ is related to the absolute error $f_{0,0}(s)-f_{0,0}^{(k)}$.  \citet{Craviotto1993Survey} make this relationship clear by furnishing an \textit{a posteriori} truncation error bound for Jacobi fractions of the same form as \eqref{eq:cfrac1} in this paper.  Assuming that $f_{0,0}^{(k)}(s) = A_k(s)/B_k(s)$ converges to $f_{0,0}(s)$ as $k\to\infty$, \citet{Craviotto1993Survey} give the bound
\begin{equation}
	\left|f_{0,0}(s) - f_{0,0}^{(k)}(s) \right| \leq \frac{\left|\frac{B_k(s)}{B_{k-1}(s)}\right|}{\left|\Im\left(\frac{B_k(s)}{B_{k-1}(s)}\right)\right|} \left| f_{0,0}^{(k)}(s) - f_{0,0}^{(k-1)}(s) \right|, 
	\label{eq:truncerr}
\end{equation}
that is valid when $\Im(s)$ is nonzero.  Note that $B_k(s)/B_{k-1}(s) = 1/D_k(s)$, so \eqref{eq:truncerr} is easy to evaluate during iteration under the Lentz algorithm.  Therefore, we stop at depth $k$ in the Lentz algorithm when
\begin{equation}
	\frac{\left|1/D_k(s)\right|}{\left|\Im\left(1/D_k(s)\right)\right|} \left| f_{0,0}^{(k)}(s) - f_{0,0}^{(k-1)}(s) \right|
	\label{eq:posterr}
\end{equation}
is small.

\subsection{Numerical results}

\begin{figure}
	\includegraphics[width=\textwidth]{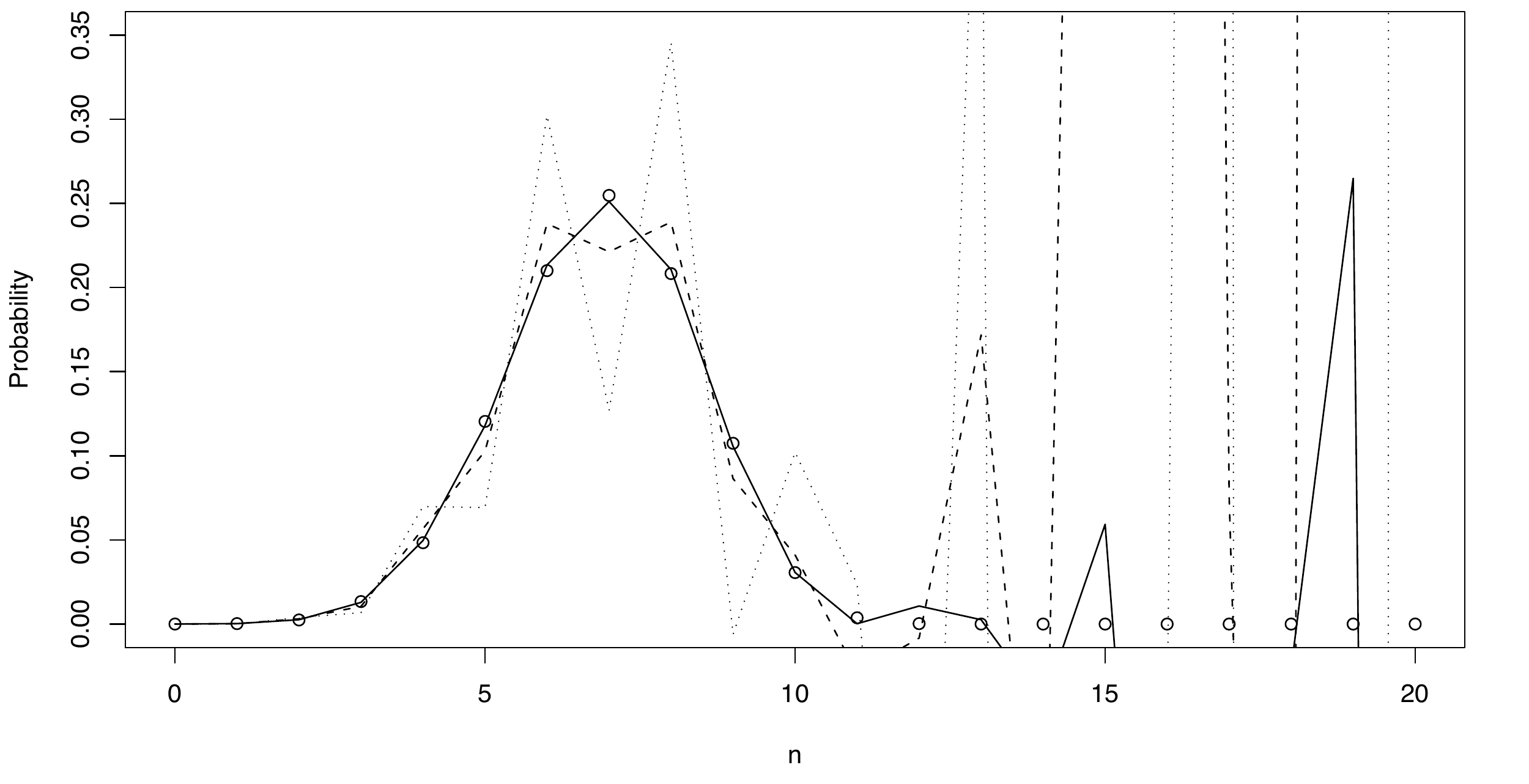}
	\caption[Comparison of transition probabilities computed by two methods]{Comparison of transition probabilities $P_{10,n}(t=1)$ computed by our error-controlled method and that of \citet{Murphy1975Some} for the immigration-death model with $\lambda_n=0.2$ and $\mu_n=0.4n$.  The open circles are the values given by our method.  The solid line corresponds with the approximant method of \citeauthor{Murphy1975Some} with $k=2$ (solid line), $k=3$ (dashed line), and $k=4$ (dotted line).  In our experience, the approximant method fails whenever $n+m+k$ is greater than approximately $20$.  It is interesting to note that increasing the depth of truncation $k$ in the approximant method actually worsens the approximation.}
	\label{fig:comparison}
\end{figure}

Although our error-controlled method is designed to be used when an analytic solution cannot be found, we seek to validate our numerical results by comparison to available analytic and numerical solutions.  For the simple BDP with $\lambda_n = n\lambda$ and $\mu_n = n\mu$, our numerical results agree with the values from the well-known closed-form solution given explicitly in \citet{Bailey1964Elements} as
\begin{equation}
	\begin{split}
		P_{m,n}(t) &= \sum_{j=0}^{\min(m,n)} \binom{m}{j} \binom{m+n-j-1}{m-1} \alpha^{m-j} \beta^{n-j} (1-\alpha-\beta)^j \\
		P_{m,0}(t) &= \alpha^m
	\end{split}
	\label{eq:simplesoln}
\end{equation}
where 
\begin{equation} 
\alpha = \frac{\mu\left(e^{(\lambda-\mu)t} - 1\right)}{\lambda e^{(\lambda-\mu)t} -\mu} \quad \text{and} \quad \beta = \frac{\lambda\left(e^{(\lambda-\mu)t} - 1\right)}{\lambda e^{(\lambda-\mu)t} -\mu}. 
\end{equation}

\citet{Murphy1975Some} give numerical probabilities for four general birth-death models: a) immigration-death with $\lambda_n=0.2$ and $\mu_n=0.4n$; b) immigration-emigration with $\lambda_n=0.3$, $\mu_0=0$, and $\mu_n=0.1$; c) queue with $\lambda_n=0.6$, $\mu_0=0$, $\mu_1=\mu_2=0.2$, $\mu_3=\mu_4=0.4$, and $\mu_n=0.6$ for $n\geq 5$; and d) $\lambda_n=0.4$, $\mu_n = 0.1\sqrt{n}$.  Our results agree with those computed by \citeauthor{Murphy1975Some} for each of the four models given in Tables 2 through 7 in their paper \citep{Murphy1975Some}.  We note that \citeauthor{Murphy1975Some} did not report probabilities for $m>2$ or $n>5$ in any of their four models.  In our experience, their method performs poorly when $n+m+k$ is greater than approximately $20$.

As a demonstration of the instability of the approximant method, we contrast the numerical results given by our error-controlled method with those obtained using the approximant method, that we implemented as described in \citet{Murphy1975Some}, except for some rescaling of intermediate quantities to avoid obvious sources of roundoff error.  Figure \ref{fig:comparison} shows this comparison, using model (a) above, for three values of the truncation index $k$.  Note that increasing the truncation depth $k$ in the approximant method does not improve the error.

\section{Applications}

Drawing on the robustness and generality of our error-controlled method, we conclude with four models in ecology, genetics, and evolution whose analytic solutions remain elusive and where past numerical approaches have fallen short.  Using our approach, computation of transition probabilities is straightforward, and the techniques outlined above may be used without modification.  Some of the examples are well-known models, and others are novel.  In some cases, the orthogonal polynomials satisfying \eqref{eq:orthogpoly} are known, and hence a solution could be numerically computed using \eqref{eq:integral}, provided there are good ways of evaluating the polynomials.  Often, a severe drawback of using known orthogonal polynomials to compute a solution based on \eqref{eq:integral} is that the polynomials are model-specific.   This makes experimentation and model selection difficult, since computation of transition probabilities depends on a priori analytic information about the polynomials and measure associated with the BDP.  Our method does not rely on a priori information about the process, other than the birth and death rates for each state.  


\subsection{Immigration and emigration}
\label{sec:imem}

\begin{figure} 
	\includegraphics[width=\textwidth]{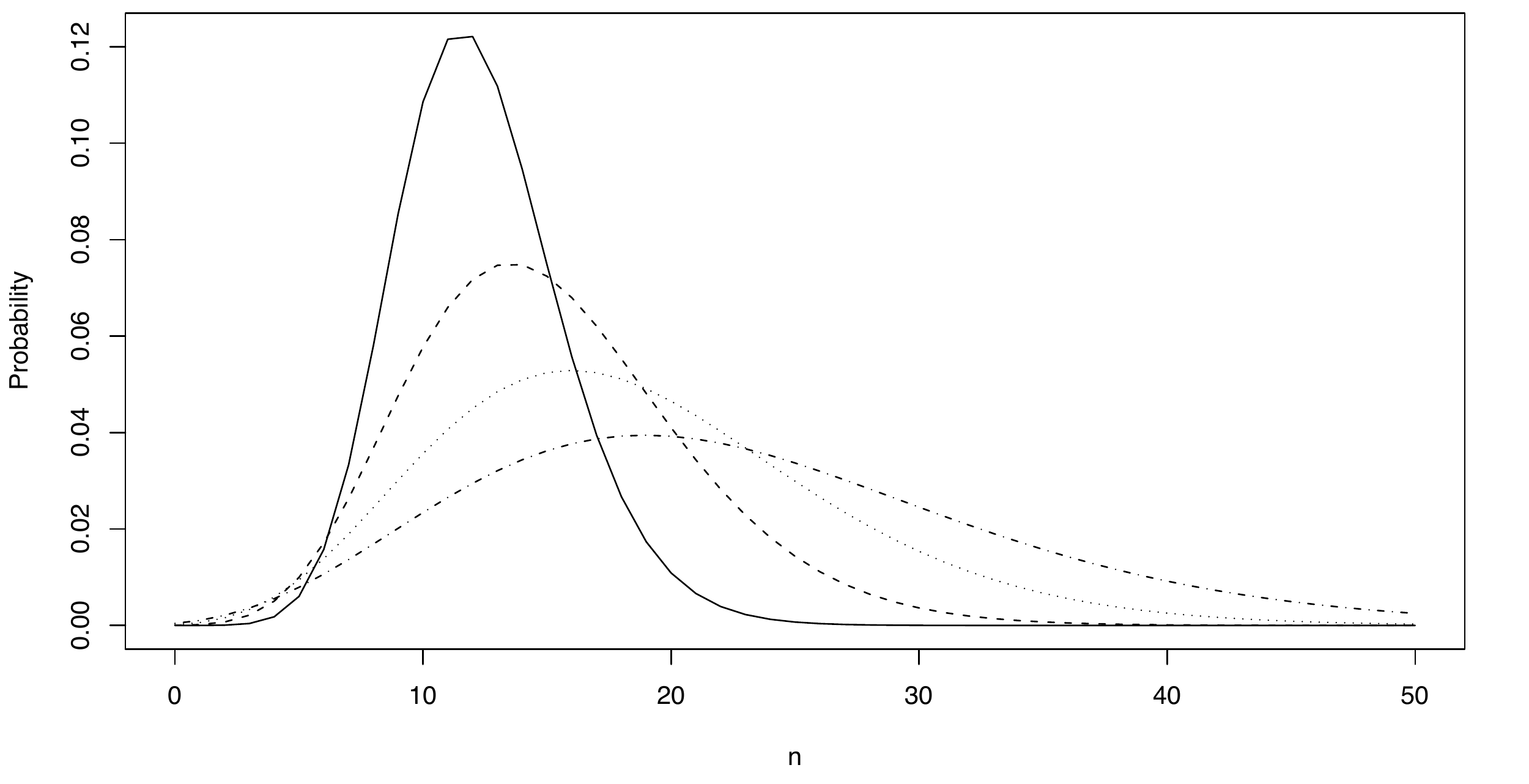} 
	\includegraphics[width=\textwidth]{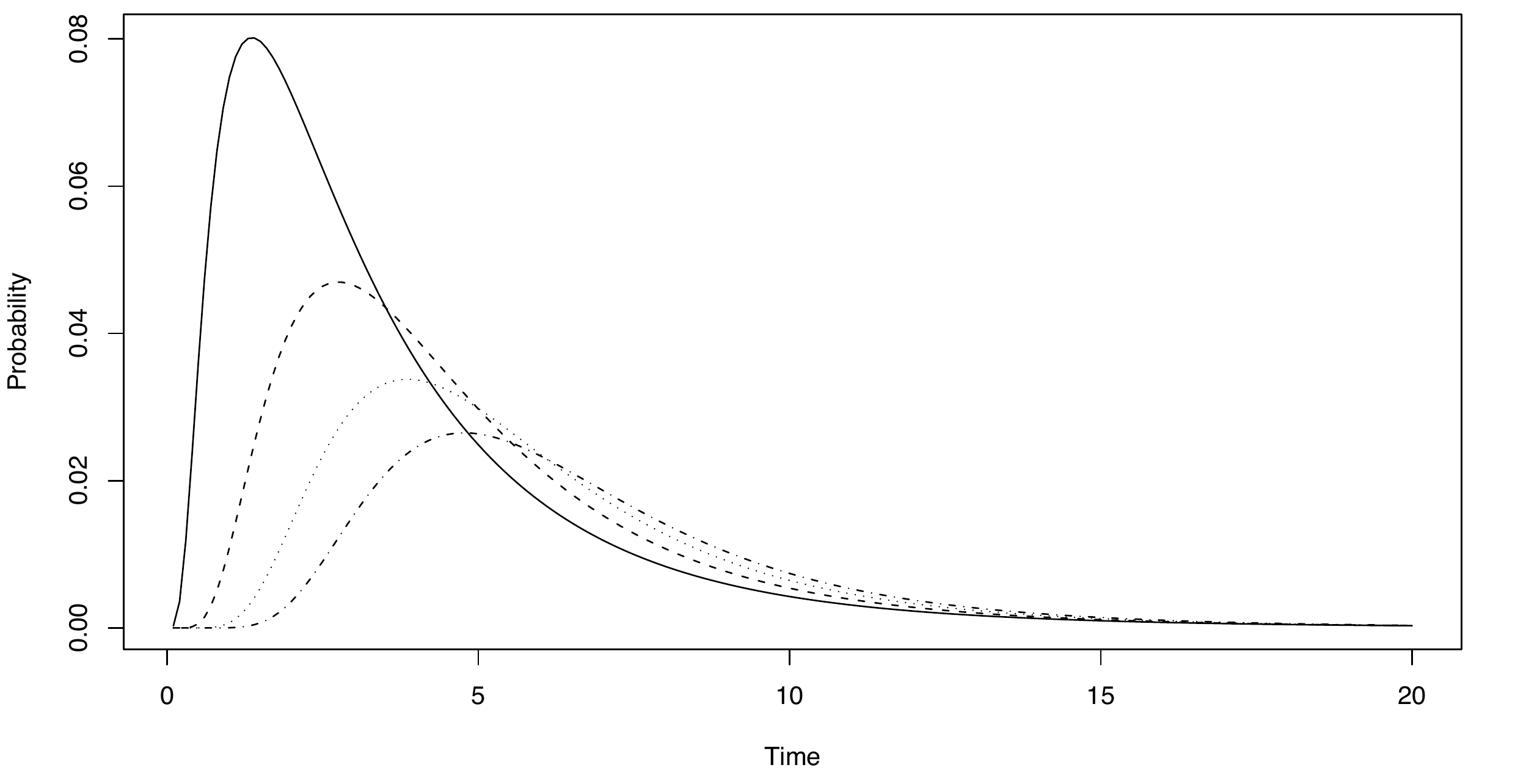} 
	\caption[Transition probabilities for the immigration/emigration model]{Transition probabilities for the immigration/emigration model with $\lambda=0.5$, $\nu=0.2$, $\mu=0.3$, and $\gamma=0.1$.  The top panel shows $P_{10,n}(t)$ with $t=1$ (solid line), $t=2$ (dashed line), $t=3$ (dotted line), and $t=4$ (dash-dotted line) for $n=0,\ldots,50$.  The bottom panel shows $P_{10,n}(t)$ with $n=15$ (solid line), $n=20$ (dashed line), $n=25$ (dotted line), and $n=30$ (dash-dotted line) for $t\in (0,20)$.  } 
	\label{fig:imem}
\end{figure}

Consider a population model for the number of organisms in an area, and suppose new immigrants arrive at rate $\nu$, and emigrants leave at rate $\gamma$.  Organisms living in the area reproduce with per-capita birth rate $\lambda$ and die with rate $\mu$.  Define the linear rates
\begin{equation} 
\lambda_n = n\lambda + \nu \quad \text{and}\quad \mu_n = n\mu + \gamma. 
\end{equation}
For the case $\gamma=0$, an analytic expression for the orthogonal polynomials is known \citep{Karlin1958Linear}.  For nonzero $\gamma$, orthogonal polynomials are available from which a solution of the form \eqref{eq:integral} may be computed \citep{Karlin1958Linear,Ismail1988Linear}.  However, using our error-controlled method, we can easily find the transition probabilities without additional analytic information.  Figure \ref{fig:imem} shows an example of the time-evolution of $P_{10,n}(t)$ for various times $t$ and states $n$, with the parameters $\lambda=0.5$, $\nu=0.2$, $\mu=0.3$, and $\gamma=0.1$.  The approximant method method of \citeauthor{Murphy1975Some} fails to produce useful probabilities for $n>10$ (not shown).


\subsection{Logistic growth with Allee effects}

\begin{figure}
\includegraphics[width=\textwidth]{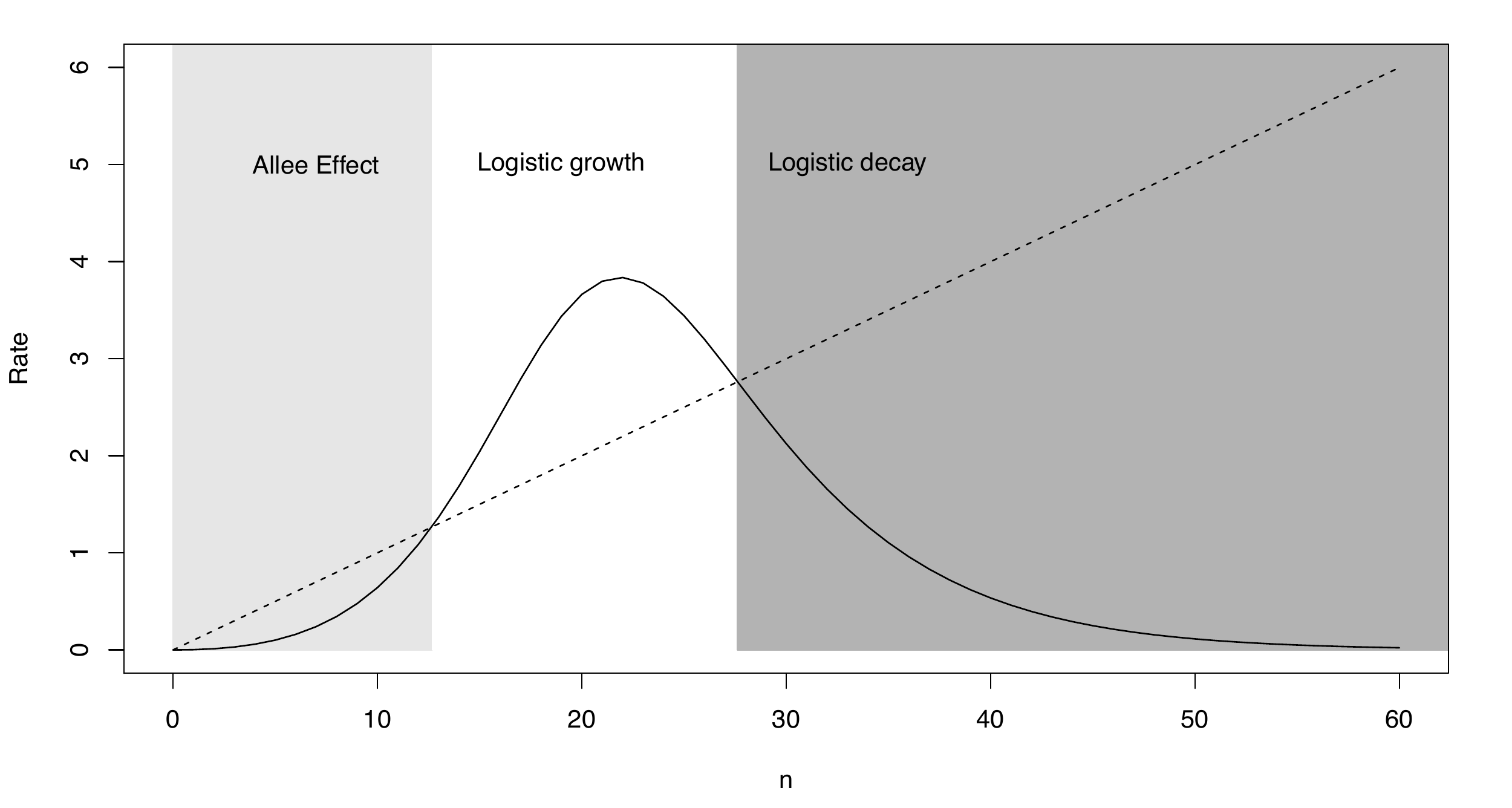}
\includegraphics[width=\textwidth]{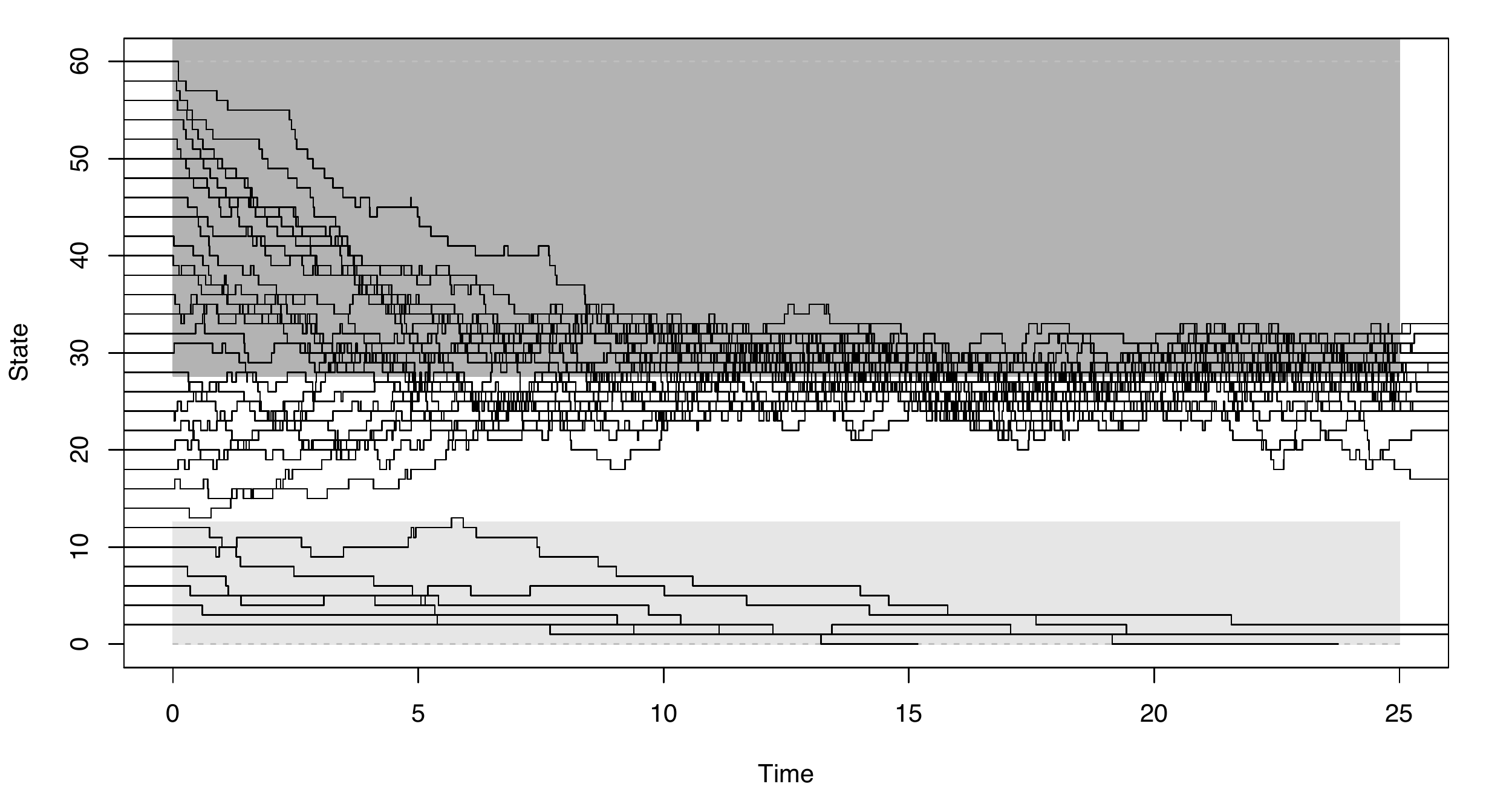}
\caption[Behavior of the logistic/Allee model]{Behavior of logistic/Allee model.  The upper panel shows a plot of birth (solid line) and death (dashed line) rates for states $n=0,\ldots,60$, and parameters $\lambda=1$, $\mu=0.1$, $M=20$, $\alpha=0.2$, and $\beta=0.3$.  The different phases of growth are labeled in the shaded regions.  The lower panel shows stochastic realizations of the logistic/Allee model for various starting values. The shaded regions correspond with the shaded phases of growth in the upper panel.}
	\label{fig:logisticrates}
\end{figure}

Populations of organisms that occupy a finite space may be subject to various constraints on their growth.  The per-capita birth rate may decline when there are more organisms than the ecosystem can sustain \citep{Tan1991Stochastic}.  This can happen when there are too many organisms competing for the same food supply.  The decay of population size above some carrying capacity is usually called logistic growth by ecologists.  Another density-dependent constraint is known as the Allee effect, in which per-capita birth rate increases superlinearly with $n$ once a small population has been established, due to favorable consequences of density, such as cooperation and mutual protection from predators \citep{Allee1949Principles}.  As a realistic example of a general BDP that has no obvious solution by orthogonal polynomials, we seek a model that both transiently supports growth above the carrying capacity, and reflects these two density-dependent constraints, similar in spirit to models described by \citet{Tan1991Stochastic} and \citet{Dennis2002Allee}.  

\begin{figure}
	\includegraphics[width=\textwidth]{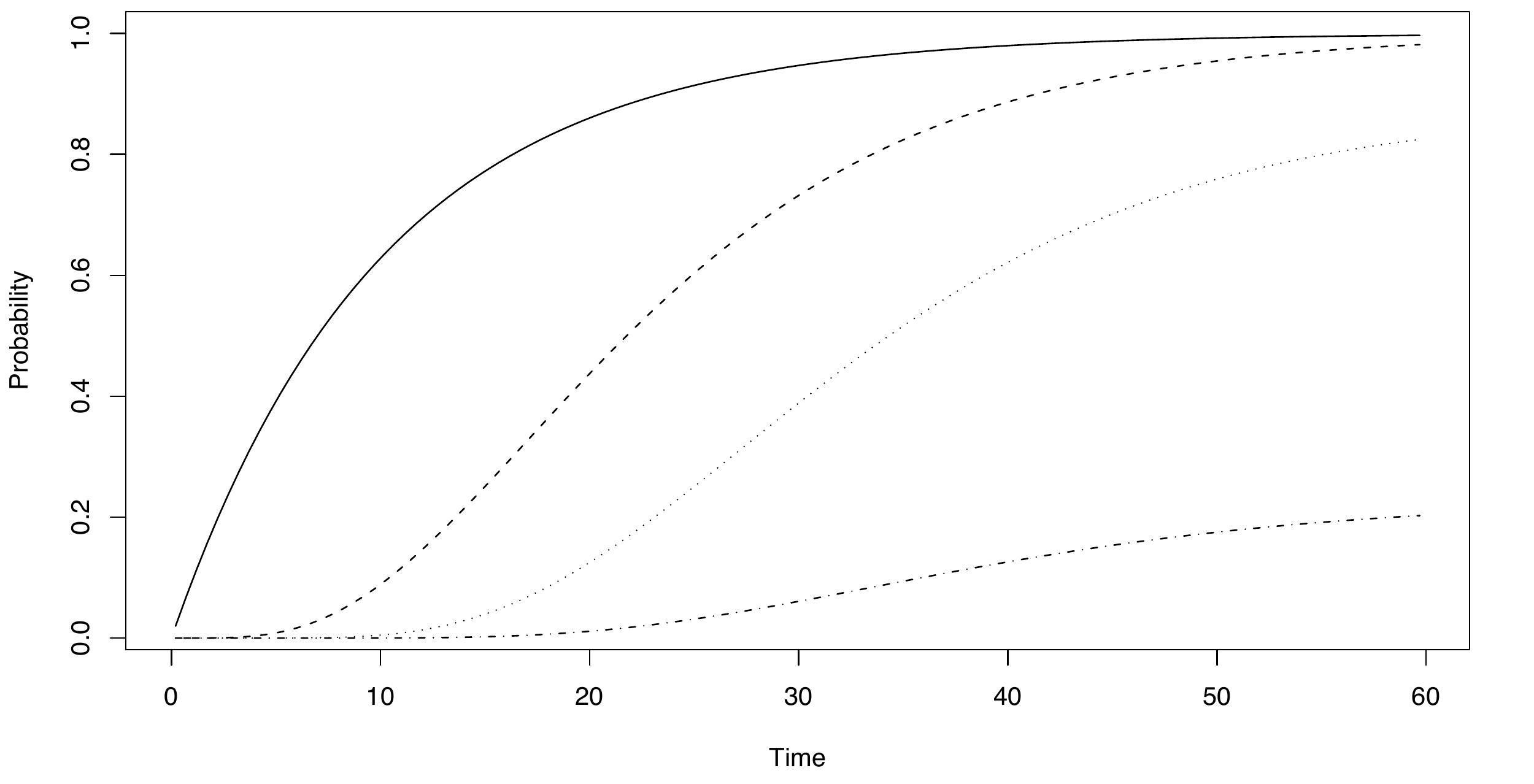}
	\caption[Extinction probabilities in the logistic/Allee model]{Logistic/Allee model probabilities of extinction $P_{m,0}(t)$ for initial population sizes $m=1$ (solid line), $m=5$ (dashed line), $m=10$ (dotted line), and $m=15$ (dash-dotted line). The full model parametrization is found in the text.}
	\label{fig:extinction}
\end{figure}

Qualitatively, if the per-capita birth rate with no density effects is $\lambda$, then the total birth rate should rise faster than $n\lambda$ when $n$ is small, slower than $n\lambda$ for intermediate $n$ near the carrying capacity, and should decay toward zero for $n$ greater than the carrying capacity.  Tan and Piantadosi introduce a logistic birth rate $\lambda_n = n\lambda\left(1-\frac{n}{N}\right)$  for a finite state space model that takes values $\{0,1,\ldots,N\}$.  However, to allow for temporary growth beyond the carrying capacity, we choose $\lambda_n \propto \lambda n^2 e^{-\alpha n}$ for intermediate and large $n$.  To achieve attenuated growth for small $n$ as well, we scale this rate by a logistic function, yielding
\begin{equation}
\lambda_n = \frac{\lambda n^2 e^{-\alpha n}}{1+e^{\beta(n-M)}} \quad\text{and}\quad \mu_n = n\mu, 
\end{equation}
where $M$ is the population size with highest birth rate, and the death rate is assumed to be proportional only to the number of existing individuals.  Figure \ref{fig:logisticrates} shows the resulting rates for various states $n$, with the different phases of population change shaded.  To illustrate that the model produces the desired behavior, several realizations of the process are given in the lower panel for various starting values.  The shaded regions correspond with the three phases of growth.  Note that most paths in the lower panel of Figure \ref{fig:logisticrates} center near $n=27$, where the birth rate and death rate are equal.  The lower panel corresponds with Figure $1$ in \citet{Dennis2002Allee}.  Figure \ref{fig:extinction} demonstrates the success of the error-controlled method in computing time-dependent extinction probabilities $P_{m,0}$ for various starting values with $\lambda=1$, $\alpha = 0.2$, $\beta=0.3$, $M=20$, and $\mu=0.1$.


\subsection{Moran models with mutation and selection}

The probability of fixation or extinction of an allele in finite populations is frequently of interest to researchers in genetics.  However, publications often rely on the probability of eventual extinction $P_{m,0}(t\to\infty)$, or the probability of fixation of a novel mutation in a population of constant size $N$, $P_{1,N}(t\to\infty)$.  While these asymptotic probabilities do reveal important properties of the underlying models, the information they provide about the distribution of time to fixation/extinction is incomplete.  In practice, researchers may observe that $m$ organisms in a sample exhibit a certain trait at a certain time.  Then $P_{m,0}(t)$, the probability of extinction of that trait at \textit{finite} times $t$ in the future should presumably be of great interest, since researchers cannot reliably observe the process for infinitely long times.  Additionally, the finite-time probability of fixation/extinction may exhibit threshold effects or unexpected dynamics that are not revealed by the asymptotic probability of such an event.

\citet{Moran1958Random} introduces a model for the time-evolution of a biallelic locus when the population size is constant through time.  A biallelic locus is a location in an organism's genome in which two different genetic variants or alleles exist in a population.  We are interested in how the number of individuals carrying each allele changes from generation to generation.  \citet{Krone1997Ancestral} exploit the Moran model to derive a BDP counting the number of individuals with a certain allele in the context of ancestral genealogy reconstruction in which one allele offers a selective advantage to individuals that carry it.  Selection greatly complicates the problem and remains an active area of research.  In a limiting case, this process corresponds to Kingman's coalescent process when there is no mutation or selection \citep{Kingman1982Coalescent,Kingman1982Genealogy}.

To construct the Moran process with mutation and selection, suppose a finite population of $N$ haploid organisms has 2 alleles at a certain locus: $A_1$ and $A_2$.  Individuals that carry $A_1$ reproduce at rate $\alpha$ and $A_2$ individuals reproduce at rate $\beta$. Suppose further that individuals carrying the $A_1$ allele have a selective advantage over individuals carrying $A_2$, so $\alpha > \beta$.  When an individual dies, it is replaced by the offspring of a random parent chosen from all $N$ individuals, including the one that dies.  This parent contributes a gamete carrying its allele that is also subject to mutation.  Mutation from $A_1$ to $A_2$ happens with probability $u$ and in reverse with rate $v$.  The new offspring receives the possibly mutated haplotype and the process continues.

Let $X(t)$ be a BDP counting the number of $A_1$ individuals on the state space $n\in\{0,\ldots,N\}$.  To construct the transition rates of the process, suppose there are currently $n$ individuals of type $A_1$.  We first consider the addition of a new individual of type $A_1$, so that $n\to n+1$.  For this to happen, the individual that dies must be of type $A_2$.  If the parent of the replacement is one of the $n$ of type $A_1$, the parent contributes its allele without mutation, and this happens with probability $1-u$.  If the parent of the replacement is one of the $N-n$ of type $A_2$, the parent contributes its allele, which then mutates with probability $v$.  Therefore, the total rate of addition is
\begin{equation}
 \lambda_n = \frac{N-n}{N}\left[\alpha\frac{n}{N}(1-u) + \beta \frac{N-n}{N} v \right],
	\label{eq:moranbirthrate}
\end{equation}
for $n=0,\ldots,N$ with $\lambda_n=0$ when $n>N$.  Likewise, the removal of an individual of type $A_1$ can happen when one of the $n$ individuals of type $A_1$ is chosen for replacement.  If the parent of the replacement is one of the $N-n$ of type $A_2$, the parent contributes $A_2$ without mutation, with probability $1-v$.  If the parent is one of the $n$ of type $A_1$, the allele must mutate to $A_2$ with probability $u$.  The total rate of removals becomes
\begin{equation}
 \mu_n = \frac{n}{N}\left[\beta\frac{N-n}{N}(1-v) + \alpha \frac{n}{N} u\right], 
	\label{eq:morandeathrate}
\end{equation}
for $n=1,\ldots,N$ with $\mu_0=\lambda_N=0$ and $\mu_n=0$ when $n>N$.  Note that if $v>0$, then $\lambda_0>0$ so the $A_1$ allele cannot go extinct.  Also, if $u>0$, then $\mu_N>0$, so the $A_1$ allele cannot be fixed in the population.

\begin{figure}
\includegraphics[width=\textwidth]{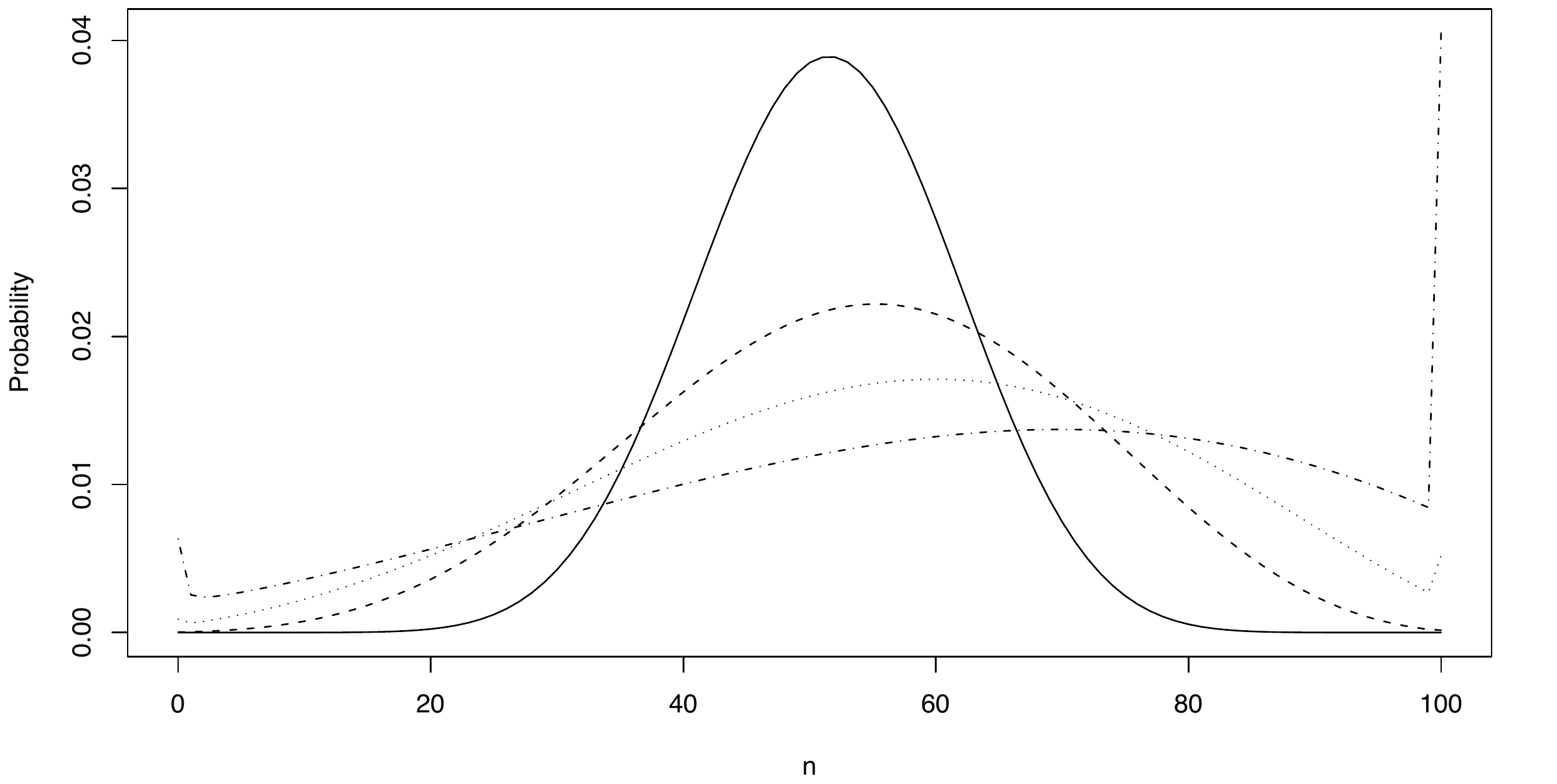}
\includegraphics[width=\textwidth]{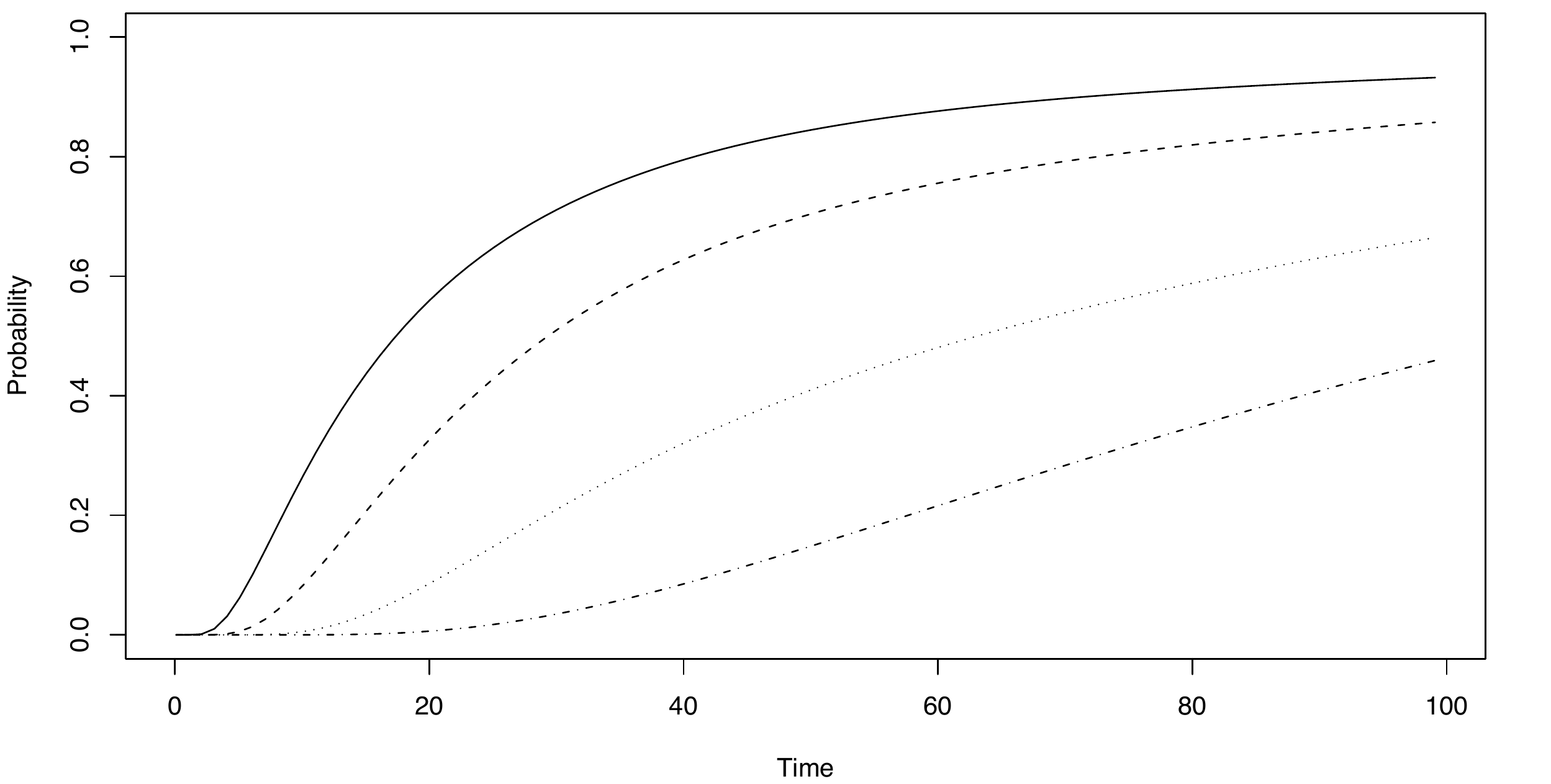}
\caption[Transition probabilities for the Moran model with selection]{Transition probabilities for the Moran model with selection.  The upper panel shows the probability of $n$ individuals having allele $A_1$ at time $t$, $P_{50,n}(t)$ for the Moran model with $N=100$, starting from $m=50$ with $u=0.02$, $v=0.01$, $\alpha=60$, and $\beta=10$.  We show the probabilities for $t=1$ (solid line), $t=3$ (dashed line), $t=5$ (dotted line), $t=8$ (dash-dotted line). Note that although the states $0$ and $100$ are not absorbing, the mutation rates $u$ and $v$ are small enough that probability accumulates significantly in these end states.  Note also the asymmetry in the distribution at longer times.  The lower panel reports the probability of fixation by time $t$, $P_{m,100}(t)$, for the same model, but with $u=0$ so the state $n=100$ is absorbing.  The probabilities shown are for $m=70$ (solid line), $m=50$ (dashed line), $m=20$ (dotted line), and $m=1$ (dash-dotted line).  Note the starkly different time-dynamics for different starting values.}
	\label{fig:moranprobs}
\end{figure}

\citet{Karlin1962Genetics} derive the relevant polynomials and measure for the Moran process described above, but without selection, so that $\alpha=\beta$.  \citet{Donnelly1984Transient} gives expressions for the transition probabilities in the case where $\alpha=\beta=1$, noting that when selection is introduced (via differing $\alpha$ and $\beta$), his approach is no longer fruitful. Using our technique, computation of the transition probabilities under selection is straightforward.  The upper panel of Figure \ref{fig:moranprobs} shows the probability of fixation by time $t$.  The lower panel shows the finite-time fixation probability of $A_1$, $P_{m,100}(t)$, with $u=0$ so the state $n=100$ is absorbing.

\begin{figure}
\includegraphics[width=\textwidth]{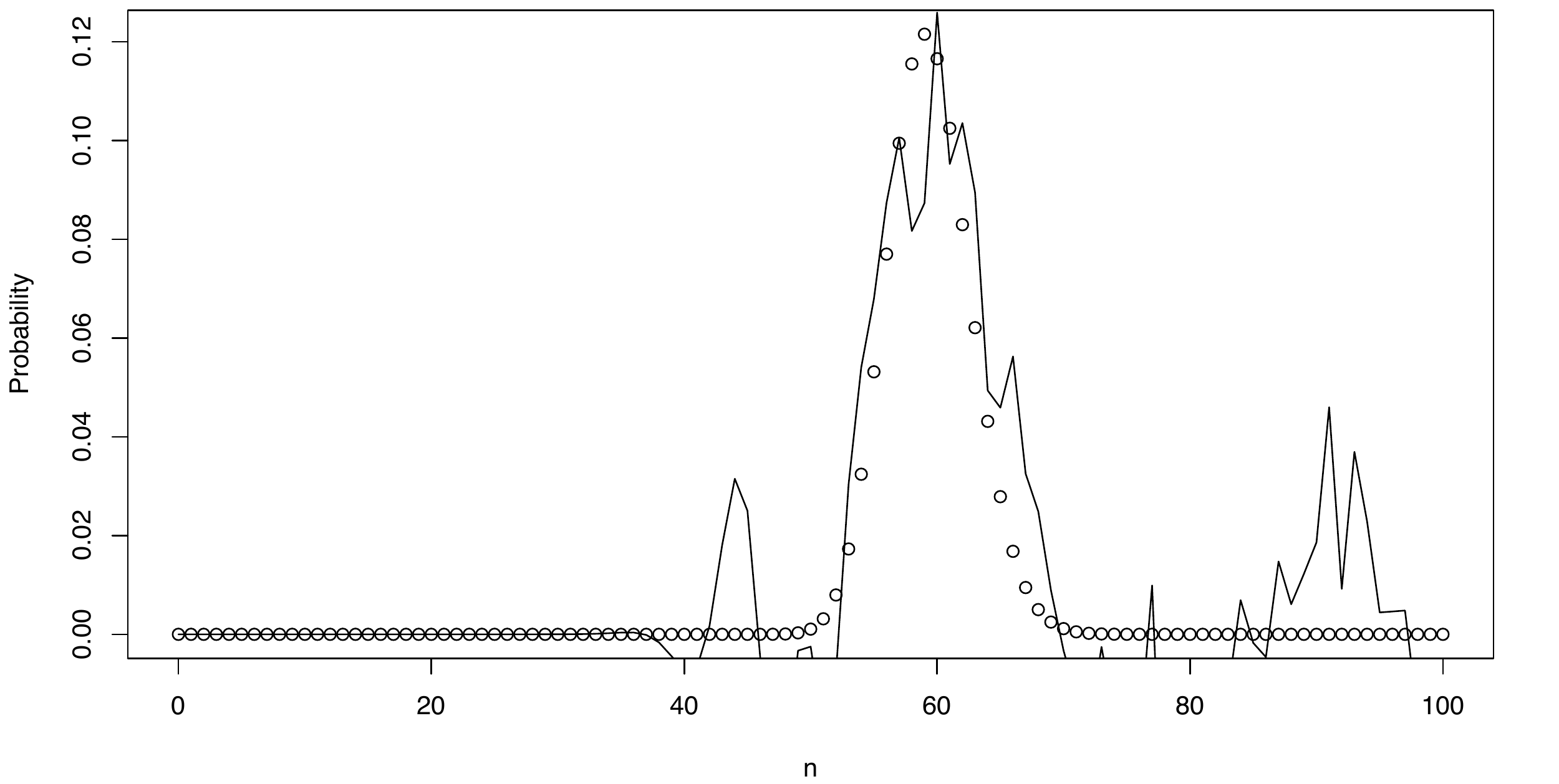}
\caption[Moran model transition probabilities]{Comparison of Moran model transition probabilities $P_{50,n}(t=0.2)$ computed by two methods with $N=100$, $\alpha=210$, $\beta=20$, $u=0.002$, and $v=0$. The open circles correspond with our error-controlled method, and the solid line corresponds with the matrix exponentiation method.  This choice of parameters causes wild fluctuations in probabilities reported by the matrix exponentiation method since the stochastic rate matrix becomes nearly singular.}
	\label{fig:eqt}
\end{figure}

Since the state space in the Moran model is finite, it is natural to consider the matrix exponentiation method discussed in the Introduction.  We write the stochastic transition matrix as
\begin{equation} 
Q = \begin{pmatrix} 
	-\lambda_0 & \lambda_0          &                    &           &        & & \\
	\mu_1      & -(\lambda_1+\mu_1) & \lambda_1          &           &         & & \\
	           & \mu_2              & -(\lambda_2+\mu_2) & \lambda_2 &        & & \\
	           &                    & \ddots             & \ddots    & \ddots & & \\
	                                &                    &           & \mu_N  & -(\lambda_N+\mu_N) & \lambda_N  
\end{pmatrix} 
\end{equation}
where $\lambda_n$ and $\mu_n$ are defined by \eqref{eq:moranbirthrate} and \eqref{eq:morandeathrate}, respectively.  In our experience, the matrix exponentiation method often works well, and its computational cost is similar to that of our error-controlled method.  However, it is highly sensitive to rate matrix conditioning.  For example, Figure \ref{fig:eqt} shows a comparison of transition probabilities from the error-controlled method and the matrix exponentiation method for the Moran model with $N=100$, $\alpha=210$, $\beta=20$, $u=0.002$, and $v=0$.  In evolutionary terms, this means that mutation from $A_1$ to $A_2$ is impossible, and the $A_2$ haplotype suffers from low fitness.  Computationally, this has the effect of making $\mu_n$ small for most $n$, and hence the rate matrix grows ill-conditioned.  

Although the rate matrix in this example is nearly defective, this choice of parameter values is not unreasonably extreme.  For example, researchers in population genetics often wish to test the hypothesis that selection occurs in a dataset.  They fit parameters for models with selection (full model) and without selection (restricted model) and perform a likelihood ratio test of this hypothesis.  If the estimates of $\beta$ and $u$ in the full model are small, they may be unable to reliably compute the probability (likelihood) of the data, given the estimated parameter values under the full model.

\subsection{A frameshift-aware indel model}

\citet{Thorne1991Evolutionary} introduce a BDP modeling insertion and deletion of nucleotides in DNA for applications in molecular evolution.  The authors model the process of sequence length evolution by assuming that a new nucleotide can be inserted adjacent to every existing nucleotide, and every existing nucleotide is subject to deletion, at a constant per-nucleotide rate.  This corresponds to the simple BDP with $\lambda_n=n\lambda$ and $\mu_n=n\mu$.  If a sequence has $m$ nucleotides at time $0$ and there are $n$ nucleotides at time $t$ later, the probability of this event is $P_{m,n}(t)$.  

\begin{figure}
\includegraphics[width=\textwidth]{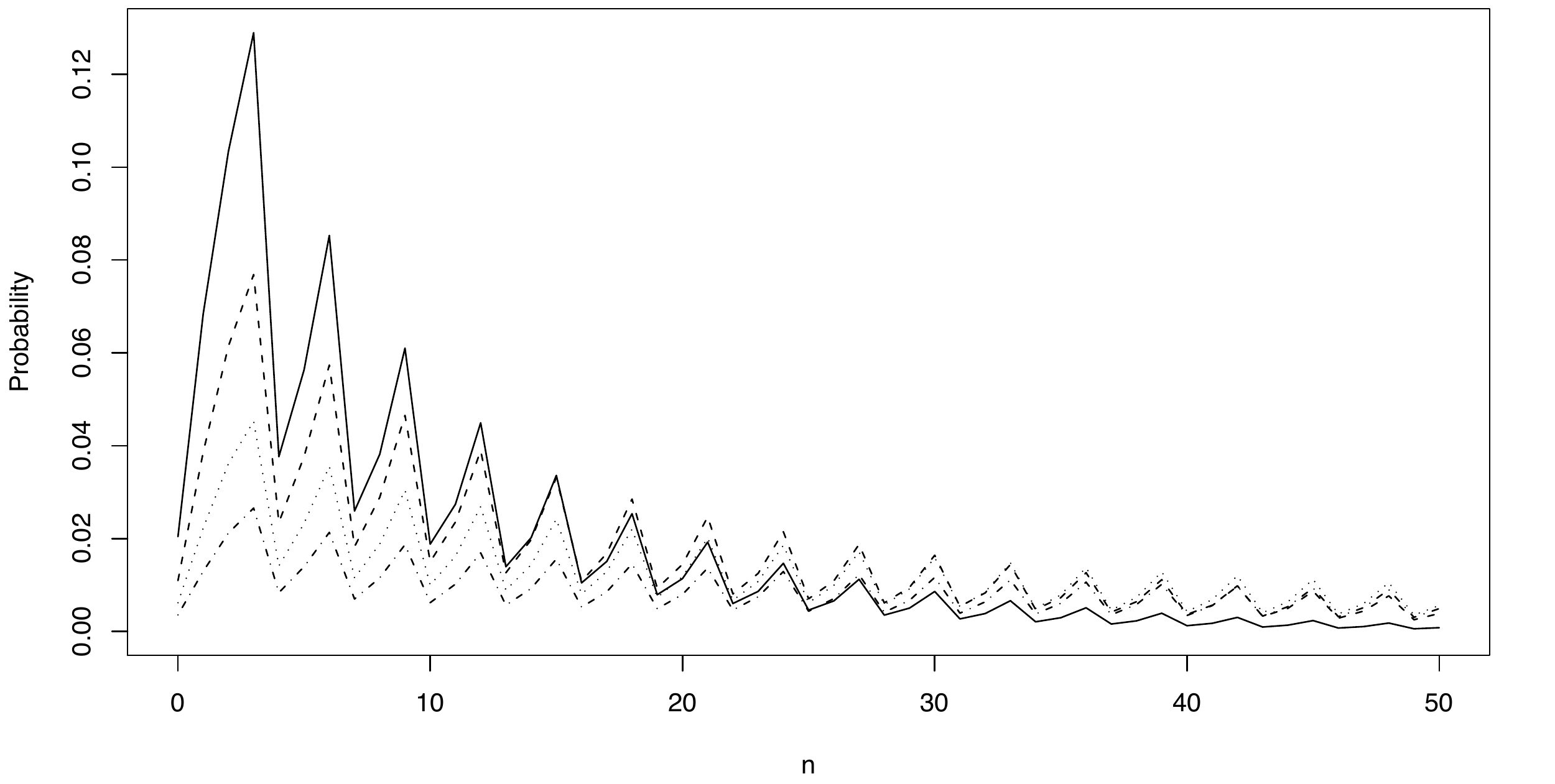}
\caption[Frameshift-aware indel model transition probabilities]{Frameshift-aware indel model probability of observing $n$ inserted DNA bases, given starting at $m=1$.  The transition probability $P_{1,n}(t)$ is shown for $t=5$ (solid line), $t=7$ (dashed line), $t=9$ (dotted line), and $t=11$ (dash-dotted line), with parameters $\beta_0 = 0.3$, $\beta_1 = 1$, $\beta_2 = 4$, $\gamma_0 = 2$, $\gamma_1 = 0.2$, and $\gamma_2 = 0.2$. }
\label{fig:indelstates}
\end{figure}

However, an important aspect of biological sequence evolution is conservation of the structure and biophysical properties of proteins that result from transcription and translation of DNA sequences.  After coding DNA is transcribed into RNA, ribosomes translate 3-nucleotide chunks (codons) of the RNA into a single amino acid residue, that is then joined to the end of a growing protein polymer.  Insertions or deletions (indels) in a DNA sequence that result in a shift in this triplet code are called ``frame-shift'' mutations.  It is likely that a frame-shift indel occurring in a protein-coding DNA sequence results in a protein that is prematurely terminated or possesses structural and chemical characteristics unlike the ancestral protein.  Insertions or deletions whose length is a multiple of three should be more common.  We seek to model this behavior in a novel way: suppose the indel process is a BDP similar in spirit to the one presented by \citet{Thorne1991Evolutionary}, and the rate of insertion and deletion of nucleotides depends on the number of nucleotides already inserted, modulo (mod) $3$:
\begin{equation}
\lambda_n = \begin{cases} 
	n\beta_0 & \text{if $n-1=0\mod 3$} \\
	n\beta_1 & \text{if $n-1=1\mod 3$} \\
	n\beta_2 & \text{if $n-1=2\mod 3$} \end{cases}
\quad\text{and}\quad 
	\mu_n = \begin{cases} 
	n\gamma_0 & \text{if $n-1=0\mod 3$} \\
	n\gamma_1 & \text{if $n-1=1\mod 3$} \\
	n\gamma_2 & \text{if $n-1=2\mod 3$} \end{cases}. 
\end{equation}
	Here we assume that $\beta_2>\beta_0,\beta_1$, and $\gamma_1>\gamma_0,\gamma_2$ so that transitions to state $n$ such that $n-1=0\mod 3$ occur at a faster rate per nucleotide.  The linear-periodic nature of these birth and death rates make solution of the orthogonal polynomials and measure corresponding with this BDP difficult.  The approximant method of \citeauthor{Murphy1975Some} also fails here for large $n$.  However, using our error-controlled method, numerical results are readily available.  Figure \ref{fig:indelstates} shows $P_{1,n}(t)$ for $n=0,\ldots,50$ at various times $t$. Note that the distribution of the number of inserted bases has peaks at the integers mod three.  Finally, it is worth noting that the dearth of tractable BDPs for indel events has been a major deterrent in statistical sequence alignment and we are actively exploring solutions to this problem using our error-controlled method.

\section{Conclusion}

Traditionally the simple BDP with linear rates has dominated modeling applications, since its transition probabilities and other quantities of interest find analytic expressions.  However, increasingly sophisticated models in ecology, genetics, and evolution, among other fields, may necessitate more advanced computational methods to handle processes whose birth and death rates do not easily yield analytic solutions.  We have demonstrated a flexible method for finding transition probabilities of general BDPs that works for arbitrary sets of birth and death rates $\{\lambda_n\}$ and $\{\mu_n\}$, and does not require additional analytic information.  This should prove useful for rapid development and testing of new models in applications.  For simple models whose solution is available, we find that our method agrees with known solutions and remains robust for large starting and ending states and long times $t$.  It is our hope that the method presented here will assist researchers in understanding the properties of increasingly rich and realistic models.


\begin{acknowledgements}
	We are grateful to Ken Lange for helpful comments.  This work was supported by National Institutes of Health grants GM086887 and T32GM008185, and National Science Foundation grant DMS0856099. A software implementation of all methods in this paper is available from FWC.
\end{acknowledgements}


\section{Appendix}

\subsection{Approximant method}
\label{sec:murphyproblems}

\citet{Murphy1975Some} approximate the inverse Laplace transform of \eqref{eq:fmnfull} by first truncating the continued fraction as a rational approximant through a partial fractions sum.  To illustrate the pitfalls of this approach, we derive the inversion expressions presented by \citeauthor{Murphy1975Some} and analyze their properties.   We provide an example to show that this technique can become numerically unstable.  We first seek to uncover the truncation error in the time domain of the transition probabilities.  If we truncate the continued fractions \eqref{eq:fmnfull} at depth $k$, we have 
\begin{equation}
	\begin{split}
		f_{m,n}^{(k)}(s) &= \left(\prod_{j=n+1}^m \mu_j\right) \frac{B_n}{B_{m+1} +} \frac{B_m a_{m+2}}{b_{m+2}+} \frac{a_{m+3}}{b_{m+3}+} \cdots \frac{a_{m+k}}{b_{m+k}} \quad \text{for $n\leq m$, and } \\
		f_{m,n}^{(k)}(s) &= \left(\prod_{j=m}^{n-1} \lambda_j\right) \frac{B_m}{B_{n+1}+} \frac{B_n a_{n+2}}{b_{n+2}+} \frac{a_{n+3}}{b_{n+3}+} \cdots \frac{a_{n+k}}{b_{n+k}} \quad \text{for $n\geq m$.} 
\end{split}
	\label{eq:fmnk}
\end{equation}
For concreteness, suppose in what follows that $n\geq m$.  Note that the denominator of the second equation is simply $B_{n+k}$. Let $A_k^{(n)}$ be the numerator of the continued fraction in the second equation in \eqref{eq:fmnk}, so 
\begin{equation}
	f_{m,n}^{(k)} = \left(\prod_{j=m}^{n-1}\lambda_j\right)\frac{A_k^{(n)}}{B_{n+k}},
	\label{eq:fmnk2}
\end{equation}
where $A_k^{(n)}$ satisfies $A_k^{(0)}=A_k$, $A_1^{(n)} = \prod_{j=1}^{n+1} a_j$, and 
\begin{equation} 
  A_k^{(n)} = a_{n+k}A_{k-2}^{(n)} + b_{n+k}A_{k-1}^{(n)}. 
  \end{equation}
Note also that the difference between truncated estimates in the Laplace domain ($s$) is
\begin{equation}
	\begin{split}
		\frac{A_{n+k}}{B_{n+k}} - \frac{A_n}{B_n} &= \frac{A_{n+k} B_n - A_n B_{n+k}}{B_{n+k} B_n} \\
		&= \frac{(-1)^n A_k^{(n)}}{B_{n+k} B_n} .
 \end{split}
 \label{eq:trunc}
\end{equation}
This yields the generalized determinant formula 
\begin{equation}
	A_{n+k} B_n - A_n B_{n+k} = (-1)^n A_k^{(n)},
	\label{eq:gendet}
\end{equation}
and at a root $s_i$ of $B_{n+k}(s)$, we have 
\begin{equation}
  A_k^{(n)}(s_i) = (-1)^n A_{n+k}(s_i)B_n(s_i).  
	\label{eq:Akn}
\end{equation}
Now if $s_1,s_2,\ldots,s_n$ are the roots of $B_n(s)$, we have, using the previous line and a partial fractions decomposition of \eqref{eq:fmnk}, the formula for the Laplace transform of the transition probability $P_{m,n}(t)$, truncated at $k$, 
\begin{equation}
	\begin{split}
		f_{m,n}^{(k)}(s) &= \left(\prod_{j=m}^{n-1}\lambda_j\right) \frac{B_m(s) A_k^{(n)}(s)}{B_{n+k}(s)} \\
		&= \left(\prod_{j=m}^{n-1}\lambda_j\right) \frac{B_m(s) A_k^{(n)}(s)}{\prod_{i=1}^{n+k} (s-s_i)} \\
		&= \left(\prod_{j=m}^{n-1}\lambda_j\right) \sum_{i=1}^{n+k} \frac{B_m(s) B_n(s_i) A_{n+k}(s_i)}{\prod_{j\neq i} (s_j-s_i)} \left(\frac{1}{s-s_i}\right), \\
	\end{split}
	\label{eq:fmnkpf}
\end{equation}
since we only require the values of $A_{n+k}(s)$ and $B_n(s)$ at the zeros of $B_{n+k}(s)$.  Then inverse transforming, an approximate formula for the transition probability $P_{m,n}(t)$ is 
\begin{equation}
	P_{m,n}^{(k)}(t) \approx \left(\prod_{j=m}^{n-1}\lambda_j\right) \sum_{i=1}^{n+k} \frac{B_m(s_i) B_n(s_i) A_{n+k}(s_i)}{\prod_{j\neq i} (s_j-s_i)} e^{-s_i t}.
	\label{eq:pnk}
\end{equation}
The roots of $B_n(s)$, used in \eqref{eq:fmnkpf} and \eqref{eq:pnk}, are often found numerically as follows. Consider the characteristic polynomial $\det\left(\tilde{B}_n + sI\right)$ of the matrix
\begin{equation}
	\tilde{B}_n = \begin{pmatrix}
   \lambda_0    &    1              &       &        &        & \\
   \lambda_0\mu_1 & \lambda_1+\mu_1 & 1     &        &        & \\    
                  & \lambda_1\mu_2 & \lambda_2+\mu_2 & 1      & \\    
                  &                 & \ddots  & \ddots & \ddots & \\ 
									&                 &   & \lambda_{n-3}\mu_{n-2}  & \lambda_{n-2}+\mu_{n-2} & 1 \\ 
									&                 &   &  & \lambda_{n-2}\mu_{n-1} & \lambda_{n-1}+\mu_{n-1} \\ 
	\end{pmatrix}.
	\label{eq:Bn}
\end{equation}
It is clear that the $n$th partial denominator $B_n(s) = \det(\tilde{B}_n + sI)$, and this quantity is zero when $-s$ is an eigenvalue of the matrix $\tilde{B}_n$. Therefore, the negatives of the eigenvalues of $\tilde{B}_n$ are the roots of $B_n(s)$.  Furthermore, $\tilde{B}_n$ can be transformed into a real symmetric matrix via a similarity transform and hence $B_n(s)$ has precisely $n$ roots, all of which are simple, real, and negative.  One usually finds these eigenvalues via the QR algorithm or similar numerical techniques \citep{Press2007Numerical}.  However, the iterative eigendecomposition of \eqref{eq:Bn} generates small errors in the eigenvalues for large $n+k$.  These errors are amplified in the product in the denominator of each summand in \eqref{eq:pnk}, resulting in a sum with both positive and negative terms that may be very large.  \citet{Klar2010Zipf} encounter similar instability in this algorithm.  Their solution is to find the roots of the terms in the numerator and compute each summand as a product of individual numerators and denominators in an attempt to keep roundoff error in the product from accumulating.  So, if $z_1,\ldots,z_{n+k}$ are the roots of $A_{n+k}$ then \eqref{eq:pnk} becomes 
\begin{equation}
	P_{m,n}^{(k)}(t) \approx \left(\prod_{j=m}^{n-1}\lambda_j\right) \sum_{i=1}^{n+k} B_m(s_i)B_n(s_i)(z_i - s_i) \prod_{j\neq i} \left(\frac{z_j-s_i}{s_j-s_i}\right) e^{-s_i t}.
	\label{eq:pnkzipf}
\end{equation}
This procedure does improve the numerical stability of the computation, but requires two eigendecompositions of possibly large matrices for every evaluation of $P_{m,n}(t)$, increasing the computational cost and, for large $m$ and $n$, the roundoff error.  In our opinion, it is more advantageous to avoid truncation of the continued fraction \eqref{eq:cfrac1} at a pre-specified index, and instead evaluate the continued fraction until convergence during numerical inversion.  Figure \ref{fig:comparison} shows how approximant methods fail for large $n$.


\subsection{A power series method}
\label{sec:parthaproblems}

\citet{Parthasarathy2005Exact} present exact solutions by transforming continued fractions such as \eqref{eq:cfrac1} into an equivalent power series.  \citet{Wall1948Analytic} shows that Jacobi fractions of this type can always be represented by an equivalent power series.  However, the small radius of convergence of power series expressions for transition probabilities can limit their usefulness for long times or large birth or death rates.  \citeauthor{Parthasarathy2005Exact} show that $P_{0,n}(t)$ has a power series representation given by
\begin{equation}
	P_{m,n}(t) = \left( \prod_{k=0}^{n-1} a_{2k} \right) \sum_{m=0}^\infty (-1)^m A(m,2n) \frac{t^{m+n}}{(m+n)!} ,
	\label{eq:pnpartha}
\end{equation}
where 
\begin{equation}
 A(m,n) = \sum_{i_1=0}^n a_{i_1} \sum_{i_2=0}^{i_1+1} a_{i_2} \sum_{i_3=0}^{i_2+1} a_{i_3} \cdots \sum_{i_m=0}^{i_{m-1}+1} a_{i_m}, 
 \end{equation}
with $A(0,n) = 1$ \citep{Parthasarathy2005Exact,Parthasarathy2006Formula}.  Here, $a_{2n} = \lambda_n$ and $a_{2n+1} = \mu_n$ in the notation used in their papers.
This approach is unique because it yields an exact analytic expression for the transition probabilities of a general BDP.  However, the radius of convergence of the power series depends on the specified rates, and this radius may be quite small.  To illustrate the pitfalls of this approach, consider $a_n = (n+1)\lambda$, corresponding to the BDP with $\lambda_n=(2n+1)\lambda$ and $\mu_n=2n\lambda$ \citep[Example 4.6]{Parthasarathy2005Exact}.  The power series for the transition probability in this process becomes 
\begin{equation}
P_{0,n}(t) = \sum_{m=0}^\infty (-1)^m \frac{(2n+2m)!}{m!n!} \frac{(\lambda t/2)^{n+m}}{(n+m)!}.
	\label{eq:parthasimple}
\end{equation}
Then the radius of convergence $R$ of the power series is given by 
\begin{equation}
	\begin{split}
		1/R &= \lim_{m\to\infty} \left| \frac{(2n+2m+2)!\left(\frac{\lambda}{2}\right)^{n+m+1}}{(m+1)!n!(n+m+1)!} \times \frac{m!n!(n+m)!}{(2n+2m)! \left(\frac{\lambda}{2}\right)^{n+m}} \right| \\
		&= \lim_{m\to\infty} \frac{(2m+2n+1)(2n+2m+2)}{(m+1)(n+m+1)}  \left(\frac{\lambda}{2}\right) \\
		&= \lim_{m\to\infty}  \frac{2m+2n+1}{m+1} \lambda \\
		&= 2\lambda  .
	\end{split}
	\label{eq:radius}
\end{equation}
And so the series diverges when $2\lambda t>1$. To illustrate the limitations of the power series approach, note that in this process, the transition intensity from $0$ to $1$ is $\lambda$, so the expected first-passage time from $0$ to $1$ is $\E(T_{0,1}) = 1/\lambda$.  Therefore, we cannot evaluate \eqref{eq:parthasimple} when $t$ is greater than $\E(T_{0,1})/2$.  If $n$ is much greater than $1$, we may be unable to reliably evaluate $P_{0,n}(t)$ for times near $\E(T_{0,n})$.


\bibliographystyle{spbasic}
\bibliography{fcrawford}

\end{document}